# Leveraging Vision-Language Models for Manufacturing Feature Recognition in CAD Designs

**Muhammad Tayyab Khan [a, c], Lequn Chen [b, *], Ye Han Ng [b], Wenhe Feng [a], Nicholas Yew Jin Tan [a], Seung Ki Moon [c, **]**

[a] Singapore Institute of Manufacturing Technology (SIMTech), Agency for Science, Technology and Research (A*STAR), 5 CleanTech Loop, #01-01 CleanTech Two Block B, Singapore 636732, Republic of Singapore

[b] Advanced Remanufacturing and Technology Centre (ARTC), Agency for Science, Technology and Research (A*STAR), 3 CleanTech Loop, #01-01 CleanTech Two, Singapore 637143, Republic of Singapore

[c] School of Mechanical and Aerospace Engineering, Nanyang Technological University, 639798, Singapore

## Abstract

Automatic feature recognition (AFR) is essential for transforming design knowledge into actionable manufacturing information. Traditional AFR methods, which rely on predefined geometric rules and large datasets, are often time-consuming and lack generalizability across various manufacturing features. To address these challenges, this study investigates vision-language models (VLMs) for automating the recognition of a wide range of manufacturing features in CAD designs without the need for extensive training datasets or predefined rules. Instead, prompt engineering techniques, such as multi-view query images, few-shot learning, sequential reasoning, and chain-of-thought, are applied to enable recognition. The approach is evaluated on a newly developed CAD dataset containing designs of varying complexity relevant to machining, additive manufacturing, sheet metal forming, molding, and casting. Five VLMs, including three closed-source models (*GPT-4o, Claude-3.5-Sonnet, and Claude-3.0-Opus*) and two open-source models (*LLava and MiniCPM*), are evaluated on this dataset with ground truth features labeled by experts. Key metrics include feature quantity accuracy, feature name matching accuracy, hallucination rate, and mean absolute error (MAE). Results show that Claude-3.5-Sonnet achieves the highest feature quantity accuracy (74%) and name matching accuracy (75%) with the lowest MAE (3.2), while GPT-4o records the lowest hallucination rate (8%). In contrast, open-source models have higher hallucination rates (>30%) and lower accuracies (<40%). This study demonstrates the potential of VLMs to automate feature recognition in CAD designs within diverse manufacturing scenarios.

## 1. Introduction

The integration of Computer-Aided Design (CAD), Computer-Aided Process Planning (CAPP), and Computer-Aided Manufacturing (CAM) is crucial for seamless operations in modern manufacturing. CAD enables the creation of digital designs that serve as the foundation for the manufacturing process [1]. CAPP acts as the bridge between CAD and CAM by analyzing the necessary manufacturing processes and conditions [2]. Subsequently, CAM uses the insights from CAPP to control and automate manufacturing equipment, transforming a CAD model into a physical product [3]. This integration not only facilitates a smooth transition from digital blueprints to physical products but also significantly improves manufacturing efficiency.

Within the integrated workflow of CAD, CAPP, and CAM, Automatic Feature Recognition (AFR) plays a pivotal role in CAPP. AFR converts CAD models into recognizable manufacturing features that CAM systems can utilize [3]. The growing complexity of CAD models, which include features related to various manufacturing processes such as machining, additive manufacturing, casting, and molding, highlights the necessity of AFR. Traditional manual methods for identifying these features are often time-consuming and prone to errors [4]. By automating this process, AFR reduces the reliance on human intervention, thereby minimizing potential variability.

AFR has a substantial impact on downstream tasks by improving the accuracy of manufacturing time and material requirement calculations, both of which are crucial for cost estimation [5]. In terms of machine tool selection, AFR helps in selecting suitable tools and fixtures. Furthermore, it can be used to optimize tool paths and reduce manufacturing time by identifying efficient strategies that minimize tool changes and repositioning. These capabilities establish AFR as a vital area of research in manufacturing automation [6].

* Corresponding authors:
E-mail addresses: chen1470@e.ntu.edu.sg (L. Chen), skmoon@ntu.edu.sg(S.K. Moon)

Despite their advantages, traditional AFR methods have significant limitations. Most existing AFR methods [7–16] are primarily focused on machining features and do not adequately address features associated with other manufacturing processes, such as additive manufacturing, sheet metal forming, casting, and molding. This limitation arises from the challenge of adapting predefined geometric patterns to recognize the diverse and complex features present in various manufacturing processes. Furthermore, traditional methods rely heavily on expert knowledge to develop and maintain feature recognition rules [12,17], which can hinder scalability. The datasets used also often lack diversity, making it difficult to manage the wide range of features encountered in real-world manufacturing scenarios. Therefore, advancing feature recognition methods to comprehensively identify features across diverse manufacturing processes is crucial for optimizing the overall manufacturing workflow.

To overcome these limitations, Vision-Language Models (VLMs) [18–20] offer a promising solution for AFR. By leveraging their ability to understand and interpret complex visual and textual information, VLMs have the potential to efficiently analyze intricate CAD images based on the input textual descriptions. This approach reduces the reliance on predefined geometric patterns and enhances the adaptability of AFR to manage the growing complexity of modern CAD designs.

To explore the potential of VLMs in manufacturing feature recognition and address the challenges posed by current AFR methods, this study aims to answer the following research questions:

1. *How effectively can VLMs recognize features in CAD designs across various manufacturing processes?*
2. *What prompt engineering techniques can be implemented to optimize VLMs for AFR tasks?*
3. *What metrics and methodologies can be used to effectively quantify and evaluate the performance of VLMs in AFR?*

The objective of this work is to automate the recognition of a wide range of manufacturing features in CAD designs using VLMs. To achieve this, a diverse dataset of 100 CAD designs is developed and categorized into easy, medium, and hard levels based on feature complexity, feature quantity, and expert judgment. The study employs four prompt engineering techniques, such as multi-view query images [21], few-shot learning [22], sequential reasoning [23], and chain-of-thought (CoT) reasoning [24,25], to enhance VLM performance in AFR tasks.

Five state-of-the-art (SOTA) VLMs are evaluated, including three closed-source models (*GPT-4o* [26], *Claude-3.5-Sonnet* [27], and *Claude-3.0-Opus* [28]) and two open-source models (*MiniCPM-Llama3-V2.5* [29] and *Llava-v1.6-mistral-7b* [30]). These models are tested against a ground truth CAD dataset labeled by domain experts. Their effectiveness is assessed using four key evaluation metrics: feature quantity accuracy, feature name matching accuracy, hallucination rate, and mean absolute error. These metrics provide a robust framework for evaluating the capabilities of VLMs to accurately identify manufacturing features across different levels of complexity.

The rest of the paper is organized as follows: Section 2 reviews related literature and outlines the challenges in AFR. Section 3 explains the methodology, including the creation of the CAD dataset and the application of prompt engineering techniques. Section 4 presents the results and discusses the performances of the VLMs. Finally, Section 5 provides the conclusions, discusses the study's limitations, and suggests directions for future research.

# 2. Related Work

In this section, we review the literature on Automatic Feature Recognition (AFR) and identify the challenges associated with its implementation. We also examine the application of VLMs in AFR and discuss the motivations driving this study.

## 2.1. Challenges in Automatic Feature Recognition

In Computer-Aided Manufacturing (CAM), features provide a higher-level abstraction of geometric shapes, which is essential for automating various Computer-Aided Engineering (CAE) activities [31]. Feature recognition involves identifying these features to seamlessly integrate design with downstream applications such as engineering analysis [32,33], reverse engineering [34], optimization [35], design validation [36], manufacturing planning [37,38], and Design for Excellence (DfX) [39]. However, CAD models often lack the high-level geometric and topological descriptions necessary for these tasks [40]. Consequently, feature recognition plays a key role in translating low-level geometric entities from CAD models into meaningful features with attributes that reflect both design intent and manufacturing functions [31]. Despite these efforts, effectively recognizing complex manufacturing features remains a significant challenge.

Over the past four decades, researchers have developed a wide range of AFR techniques, broadly classified into rule-based and learning-based approaches [6–8,17,41]. Early efforts focused on rule-based approaches, which rely on expert-defined rules to identify machining features based on their geometric and topological properties [17,42]. Rule-based techniques include logic rule and expert systems, volume decomposition, hint-based methods, graph-based methods, and hybrid systems [12,31]. Logic rules utilize expert-defined rules to recognize features based on specific geometric patterns, offering straightforward implementation but lacking in scalability and flexibility [17,43,44]. Graph-based methods model CAD designs using nodes and arcs to represent faces and edges, enhancing feature recognition performance but facing challenges with topology variations and high computational demands [45–49]. Hint-based approaches use retained data from machining features to infer types yet struggle with the flexibility needed for complex features [50,51]. Volume decomposition, which divides a CAD model into smaller volumes for feature assignment, faces similar issues with computational intensity [52–54]. Despite the integration of these methods into hybrid approaches to address complex feature recognition, they continue to confront significant challenges in managing intersecting features and computational efficiency, highlighting the need for more adaptable solutions in AFR [55].

In recent years, learning-based methods reduce reliance on extensive design rules through machine learning (ML) and deep learning (DL) techniques [7,9]. These approaches are categorized into Artificial Neural Network (ANN)-based and Convolutional Neural Network (CNN)-based methods, which convert machining features into low-level representations using extensive training datasets [12]. ANNs utilize face adjacency matrices or vectors that represent geometry and topology information as inputs [56,57]. However, these methods face challenges such as dependency on complex vector design and the loss of crucial part details, which reduce recognition performance and limit the range of identifiable features [8,31]. To overcome these limitations, CNNs have been introduced with a two-step process: first, features are segmented from part models using algorithms like watershed or hint-based techniques; then, they are identified using 2D or 3D CNNs [58–60]. Recent advancements have simplified this to a one-step approach, using 2D images from various orientations to represent 3D models and predict cuboid bounding boxes that locate and identify features [61]. However, this method still struggles with challenges such as loss of geometry and topology information and inaccuracies in feature segmentation, impacting crucial downstream process planning tasks [12].

The limitations of traditional AFR methods highlight the need for more adaptable solutions. VLMs offer a promising alternative by combining computer vision with natural language processing (NLP) to interpret complex visual and textual information [18]. Unlike rule-based AFR methods that rely on predefined geometric rules, VLMs are flexible enough to recognize a variety of intricate manufacturing features. Furthermore, VLMs differ from learning-based methods, which typically require extensive training datasets, as they can effectively identify features through prompt engineering techniques [19,23,25]. This strategy enables precise VLM guidance without large-scale data requirements. While traditional AFR methods often focus on specific stock geometries like cuboids [7,10,12] or cylinders [62,63], VLMs handle a broader range of stock types and features input images. Their capability to process both visual and textual data allows them to recognize diverse features in CAD designs, overcoming the scalability challenges faced by existing AFR techniques. By improving the adaptability of feature recognition, VLMs enhance critical downstream tasks such as tool selection, process planning, and cost estimation.

## 2.2. Large Language Models and Vision-Language Models for AFR

Large Language Models (LLMs) are powerful computational models designed to understand and generate human language [64–67]. LLMs like *GPT-4* [26] and *LLaMA* [68], built on Transformer architectures, have revolutionized natural language processing by enabling efficient sequential data handling and capturing long-range text dependencies [69–71]. These models excel in in-context learning (ICL), allowing them to acquire skills and adapt quickly to new tasks without retraining, making them invaluable for a variety of applications [22,72]. Similarly, conversational interfaces such as Google's *Bard* [73] and OpenAI's *ChatGPT* [26] have simplified access to complex information, enabling users to address tasks from medical diagnostics to manufacturing design without the need for complex technical jargon [20,74]. However, tasks requiring spatial understanding highlight the limitations of text-only interfaces. In such cases, visual representations are essential, as seen in manufacturing where engineers use CAD drawings to accurately communicate intricate details of design and assembly [75,76].

To address these challenges, VLMs have been developed to process both image and natural language text. Significant progress has been made in leveraging VLMs across various applications, including image manipulation (e.g., *StyleCLIP* [77], *DiffusionCLIP* [78]), text-based video retrieval (e.g., *X-CLIP* [79]) and manipulation (e.g., *Text2Live* [80]), as well as 3D shape and texture manipulation (e.g., *AvatarCLIP* [81], *CLIPFace* [82], *Text2Mesh* [83]). More recently, advanced multimodal language models such as *GPT-4o* [26] and *Claude-3.5-Sonnet* [27] have shown significant potential since their launch. These models are capable of processing both images and text inputs, producing text outputs. Given the visual nature of manufacturing features in CAD designs, leveraging the capabilities of such models offers an effective approach to overcome AFR challenges.

Traditional AFR methods heavily depend on expert-defined rules or extensive training data and face difficulties with 3D geometric details. In contrast, VLMs may offer a more adaptable solution for recognizing complex features through prompt engineering, which can reduce the dependence on large datasets and minimize the need for human intervention.

The pioneering study by Picard et al. [20] demonstrated the potential of VLMs, particularly OpenAI's *GPT-4V* [84], in AFR. Their research utilized a dataset comprising 20 CAD designs with common machining features, marking a significant initial exploration into VLM applications for AFR. However, the dataset's limited scope did not fully capture the complexity of diverse manufacturing processes. Moreover, the evaluation metrics employed were not comprehensive enough to thoroughly assess VLMs' capabilities in recognizing a wide range of manufacturing features. More in-depth study is needed to incorporate a broader range of CAD designs and develop more rigorous evaluation methods to better understand and enhance the capabilities of VLMs in AFR.

To this end, this study addresses AFR challenges and aims to enhance accuracy by analyzing five SOTA VLMs using prompt engineering strategies. The research employs a diverse CAD dataset, each CAD model varying in complexity based on feature complexity, feature quantity, and expert judgement. Four key evaluation metrics are defined to systematically analyze the VLMs' performance. This systematic approach is intended to advance the capabilities of VLMs in AFR, making them more applicable to the manufacturing domain.

# 3. Methodology

This section describes the proposed methodology, detailing the creation of a custom CAD dataset, the application of VLMs, the use of prompt engineering strategies, and the definitions of the proposed evaluation metrics.

Figure 1 outlines the proposed methodology for evaluating VLMs in AFR from CAD designs. The process starts with converting a CAD file into three different image views using the *CAD2Image* converter, based on the Python Open Cascade (*PyOCC*) library [85]. These images, combined with proposed prompts outlining the analysis criteria, serve as inputs to five SOTA VLMs. A series of six distinct experiments based on the proposed prompt engineering techniques are conducted, instructing the VLMs to adhere strictly to the provided prompts and output format. The VLMs generate answers in JSON format, listing identified features from a predefined list of manufacturing features, including details about each feature's existence and quantity. The VLMs' performance is then evaluated using four key metrics, comparing the identified features to ground truth features to ensure both accuracy and adherence to the evaluation criteria.

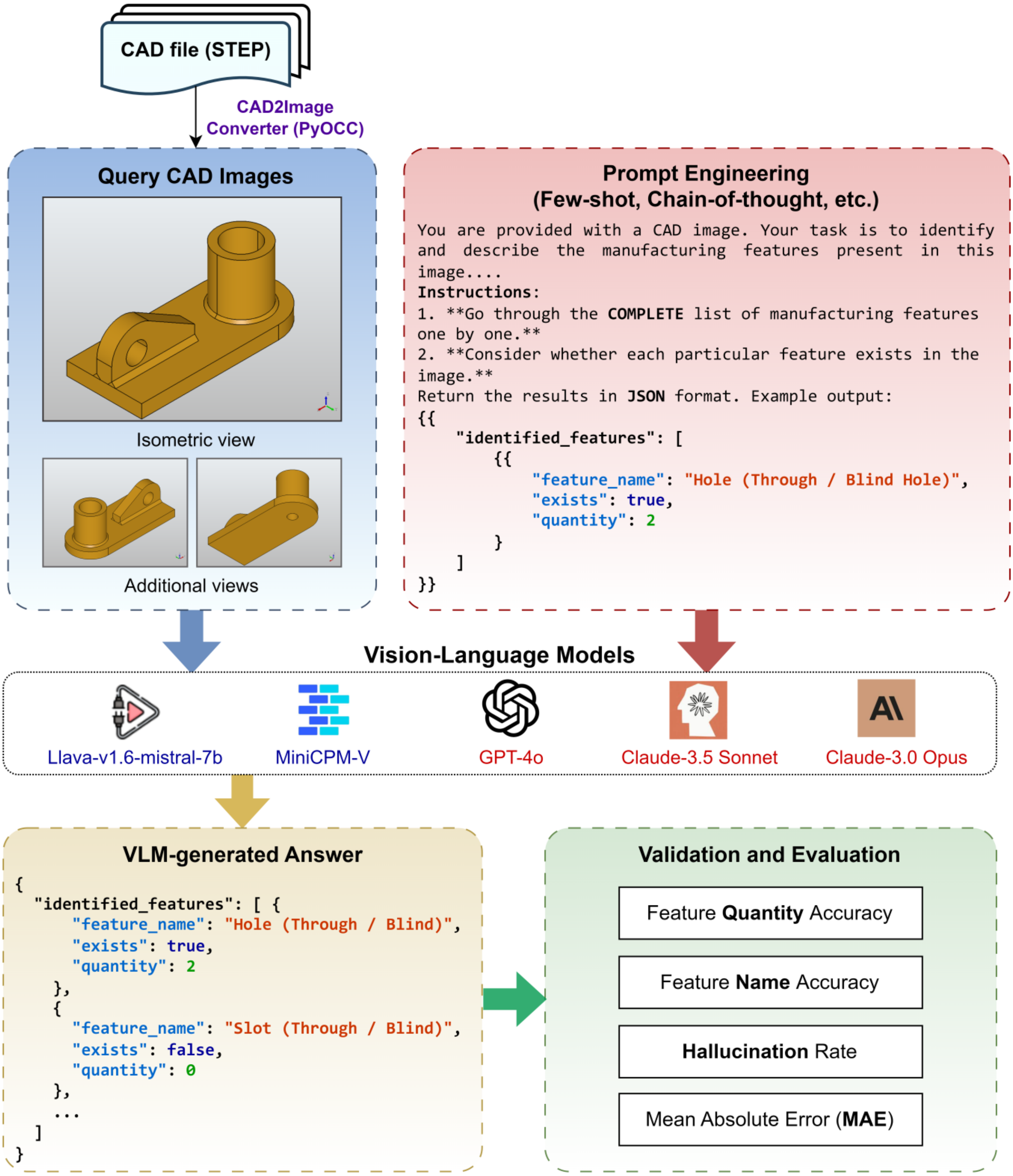

**Fig. 1** **Methodology overview.**

## 3.1. Dataset Creation

To explore the capabilities of VLMs in AFR, the study starts with the creation of a customized dataset consisting of 100 unique CAD designs, each annotated with a detailed list of manufacturing features. The dataset includes a broad range of features pertinent to various manufacturing processes and is organized into primary categories and subcategories to facilitate in-depth analysis. The main categories include *machining features*, *extrusion features*, *freeform features*, *molding and casting features*, and *sheet metal features*. Machining features are further divided into subcategories such as holes, slots, steps, pockets, edges and contours (e.g., chamfers, fillets), threads and spirals, and additional features like necks. Extrusion features include pipes, tubes, and various boss shapes. Freeform features consist of depressions and protrusions, while molding and casting features comprise ribs, gussets, and drafts. Sheet metal features encompass various features related to sheet metal fabrication such as bending, punching, stamping, etc.

The dataset is categorized into three levels of complexity: easy, medium, and hard, determined by feature quantity, complexity and human expert judgment. The easy-level dataset includes 33 CAD parts, each featuring fewer than six simpler features and excluding complex features such as gussets, ribs, necks, threaded features, and sheet metal features. The medium-level dataset contains 33 CAD parts with designs of moderate complexity and includes non-traditional cylindrical stocks with a limited presence of complex features like sheet metal bends and threaded components. The difficult-level dataset consists of 34 CAD models, characterized by highly complex designs with numerous features, including intricate geometries such as casting and freeform features.

Figure 2 provides visual examples from each dataset category, illustrating the varying levels of complexity. The easy-level examples showcase parts with fewer and simpler features, while the medium and difficult levels display increasingly complex designs. The CAD files are primarily designed by the authors, while particularly challenging parts are sourced from *GrabCAD* [86]. Each part is converted into three different image views using the *PyOCC* library to serve as input for the VLMs.

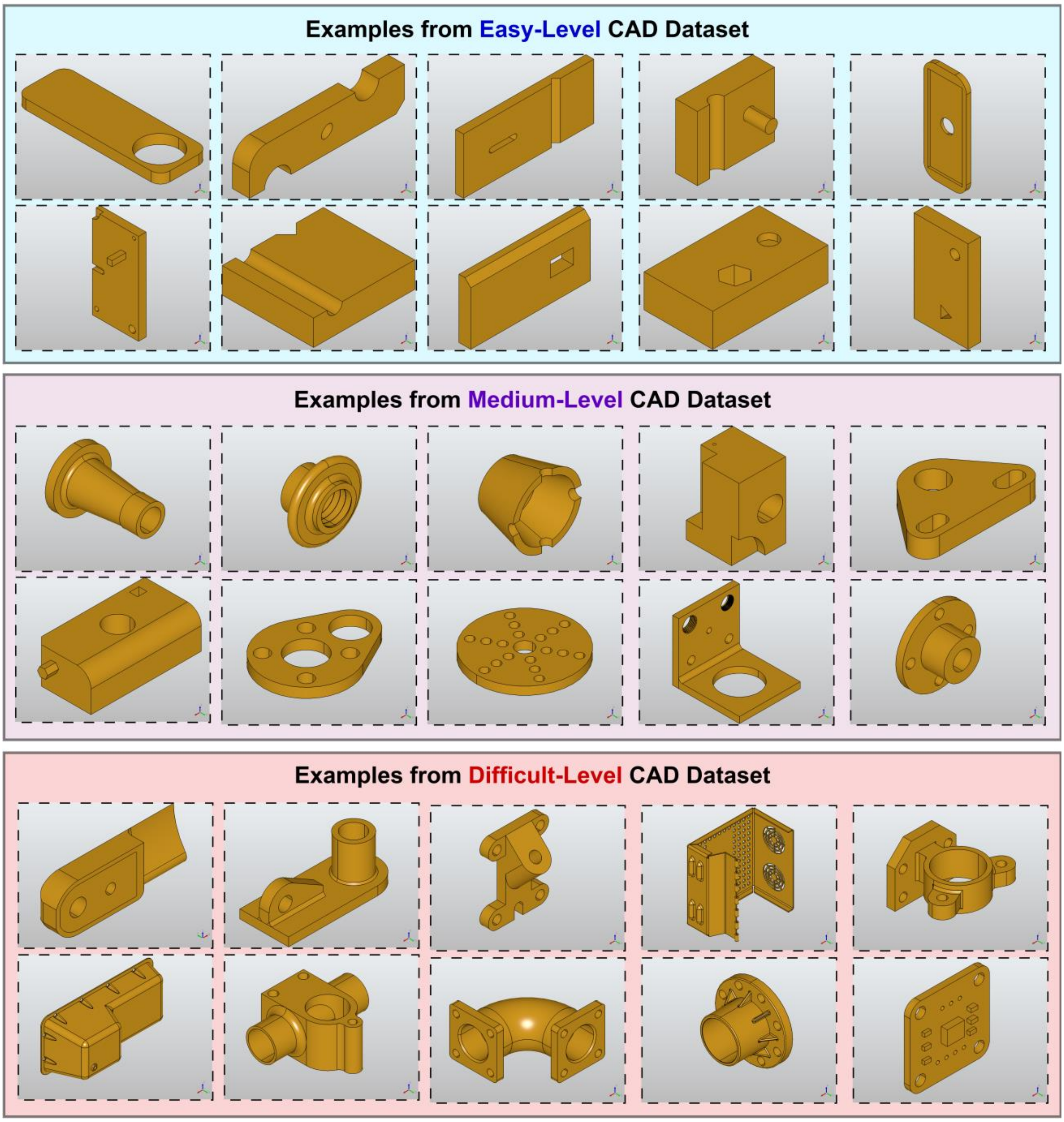

**Fig. 2 Example CAD images from each category.**

Each CAD part is manually labelled by domain experts, specifying the presence and quantity of each feature. This ground truth data is crucial for evaluating the performance of the VLMs. Figure 3 illustrates the feature distribution across the dataset's different complexity levels, highlighting variations such as the high prevalence of the '*hole*' feature in more complex CAD designs. Additionally, the hard category contains a notably higher occurrence of intricate features like sheet metal and freeform features, which are less common or absent in the easy and medium categories.

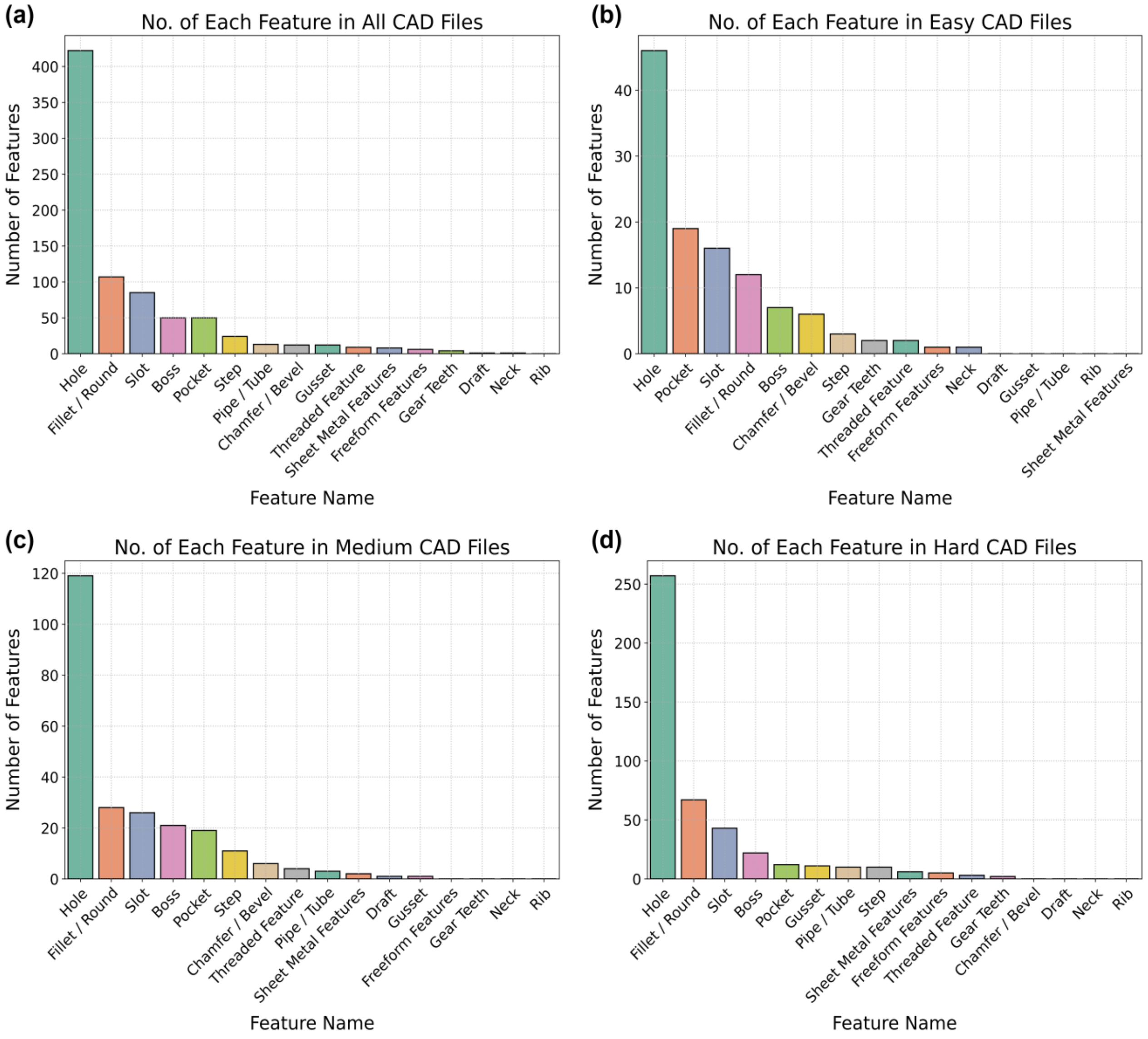

**Fig. 3 Feature distribution across different dataset categories.**

## 3.2. Vision-Language Models (VLMs)

In this study, five SOTA VLMs including three closed-source models, and two open-source models are implemented for manufacturing feature recognition:

- OpenAI's *GPT-4 omni* (`gpt-4o`) [87] is selected, which is a closed-source VLM. GPT-4o is OpenAI's most advanced, cost-efficient, multilingual, and multimodal model that integrates visual inputs with text using a transformer-based architecture. It excels in traditional benchmarks for text, reasoning, and coding intelligence, while setting new records for multilingual, audio, and vision capabilities [88]. Its affordability and high accuracy in real-time vision tasks make it particularly well-suited for recognizing diverse manufacturing features in CAD models.

- Anthropic's *Claude-3.5-Sonnet* (`claude-3-5-sonnet-20240620`) [89] is a closed-source VLM that excels in vision benchmarking data, demonstrating exceptional performance in interpreting charts, graphs, and transcribing text from imperfect images [27].

- Anthropic's *Claude-3-Opus* (`claude-3-opus-20240229`) [89] is a closed-source VLM known for its advanced multimodal capabilities within the Claude 3 series. It demonstrates near-human comprehension and excels in complex vision tasks [28].
- *MiniCPM-Llama3-V2.5 (*`MiniCPM-Llama3-V-2_5`) [29] is an open-source VLM that matches GPT-4V-level performance in a more compact form. It is optimized for efficient deployment on mobile and edge devices and surpasses similar VLMs in its class on benchmark datasets [90].
- *Llava-v1.6-mistral-7b* (`llava-v1.6-mistral-7b-hf`) [30] is an open-source VLM that offers enhanced image processing and logical reasoning capabilities. Built on the Mistral-7B language model, it achieves competitive performance in high-resolution multimodal data tasks [91].

For all experiments described in the subsequent sections, the temperature for each VLM is set to zero.

## 3.3. Prompt Engineering

Prompt engineering involves creating natural language instructions, or prompts, to systematically extract knowledge from large language models (LLMs). Unlike traditional models, which often require extensive parameter re-training or fine-tuning for specific natural language processing (NLP) tasks, the prompt engineering leverages the embedded knowledge within LLMs. This approach eliminates the need for extensive re-training, allowing researchers to interact with LLMs using natural language to achieve specific NLP tasks [25].

To address the second research question on enhancing the performance of VLMs, four commonly used prompt engineering techniques [23] are implemented in this study: multi-view query images, few-shot learning, sequential reasoning, and chain-of-thought (CoT) reasoning. These prompt engineering techniques are discussed in detail in this section.

To assess the effectiveness of these strategies, we conduct six distinct experiments using various combinations of the prompt engineering techniques on each VLM:

- *Experiment 1: Zero-shot learning with a single CAD image view processed in parallel.*
- *Experiment 2: Zero-shot learning with a single CAD image view processed sequentially.*
- *Experiment 3: Zero-shot learning with multiple CAD image views.*
- *Experiment 4: Multiple image views with few-shot learning.*
- *Experiment 5: Multiple image views with zero-shot learning and CoT reasoning.*
- *Experiment 6: Multiple image views with few-shot learning and CoT reasoning.*

For all experiments, the temperature hyperparameter is set to zero to ensure consistent response behavior. Each experiment begins with a preamble to provide the VLMs with context, followed by the introduction of a manufacturing feature list and criteria under evaluation. Illustrations of these experiments are provided in Appendix A (Fig. A1–A6).

### 3.3.1. Manufacturing Feature List

In this study, a comprehensive dataset of CAD models is created, featuring a diverse range of manufacturing features. These features are organized hierarchically to facilitate the recognition and categorization tasks performed by VLMs. Figure 4 illustrates this hierarchical organization in JSON format, where features are classified into five primary categories: machining features, extrusion features, freeform features, molding & casting features, and sheet metal features.

Each primary category is subdivided into relevant subcategories. For instance, within the *Edges & Contours* subcategory, features such as '*chamfer*' and '*fillet*' are included, while the *Threads & Spirals* subcategory lists features like '*threads*' and '*gear teeth*'. These individual features, positioned at the lowest level of the hierarchy, are the specific targets for VLM recognition in CAD parts. The dataset contains 16 such features in total, and each CAD part is labelled with these ground truths.

This hierarchical structure categorizes features with similar geometric characteristics and manufacturing processes for efficiency. For example, various types of holes (such as blind, through, countersink, and tapered, etc.) are grouped under *Holes* due to their similar geometries and tooling requirements. This approach not only streamlines the VLMs' evaluation process by making features easily accessible but also ensures the input is concise to avoid overloading the VLMs.

**Manufacturing Feature List**

```
{
    "Manufacturing Features": {
        "Machining Features": {
            "Hole (Through / Blind Hole)": [],
            "Slot (Through / Blind / T-Slot / Dovetail)": [],
            "Step (Through / Blind Step)": [],
            "Pocket (Blind / Through / Circular End Pocket)": [],
            "Edges & Contours": {
                "Chamfer / Bevel (Sharp Edge)": [],
                "Fillet / Round (Concave / Convex)": []
            },
            "Threads & Spirals": {
                "Threaded Feature": [],
                "Gear Teeth": []
            },
            "Additional Machining Features": {
                "Neck": []
            }
        },
        "Extrusion Features": {
            "Pipe / Tube": [],
            "Boss (Circular / Obround / Irregular / Rectangular, etc)": []
        },
        "Freeform Features (Depression, Protrusion)": [],
        "Molding & Casting Features": {
            "Rib": [],
            "Gusset": [],
            "Draft": []
        },
        "Sheet Metal Features": []
    }
}
```

**Fig. 4 Manufacturing feature list.**

### 3.3.2. Single View vs. Multi-view Query Images

Building on the hierarchical feature organization discussed earlier, this section explores how single view and multi-view query images are used to potentially improve the performance of VLMs in AFR. As illustrated in Fig. 5, the preamble for each type of image provides specific prompts to the VLM, instructing it to identify and describe the manufacturing features from the images. These prompts are crafted using the feature names from the comprehensive list outlined in previous section.

**Preamble (Single-view query image)**

```
You are provided with a CAD image. Your task is to identify and describe the manufacturing features present in this image. Use the provided list of feature names, which is delimited by triple backticks. Use this list as a reference and strictly follow the naming conventions:

'''{manufacturing_features_names}'''
```

**Preamble (Multi-vew query image)**

```
You are provided with images containing multiple views of a 3D CAD model taken from different angles. Your task is to identify and describe the manufacturing features present in these images. Use the provided list of feature names, which is delimited by triple backticks. Use this list as a reference and strictly follow the naming conventions:

'''{manufacturing_features_names}'''
```

**Fig. 5 Preamble prompt.**

For single view query images, a single CAD image is provided to the VLM. This setup restricts the VLM to analyzing the object from only one angle, limiting the perspective available. Conversely, multi-view query images, as shown in Fig. 6, present three different views of a CAD model. This approach offers a broader visualization, potentially allowing the VLMs to access features that are not visible from a single isometric view. We anticipate that this method will result in higher accuracy, as incorporating multiple views significantly enhances the visual spatial reasoning capabilities of VLMs [92].

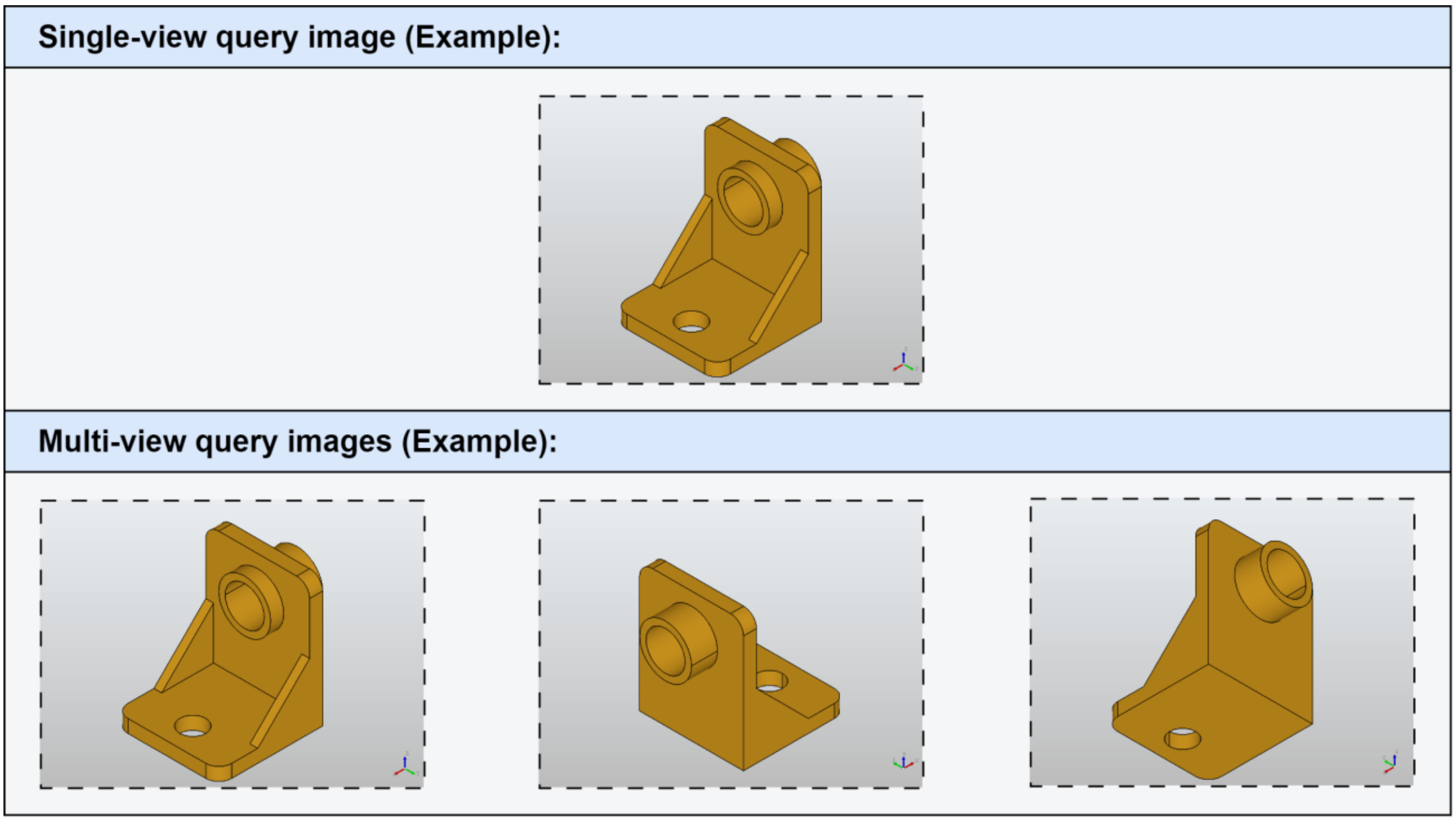

**Fig. 6 Query CAD images: single view and multi-view.**

### 3.3.3. Zero-shot vs. Few-shot Prompt

Zero-shot and few-shot prompts are techniques used to interact with LLMs and VLMs across various domains. Zero-shot prompting requires language models to perform tasks based solely on their pre-trained knowledge, without prior examples [72]. In this study, VLMs are directed to identify manufacturing features using an input query image along with textual prompts in a zero-shot learning approach, without the benefit of additional labelled examples.

Conversely, few-shot prompting provides the language models with one or more examples to impart task-specific insights. This approach generally outperforms zero-shot prompting in various benchmarks [22]. In this study, a *one-shot learning* method [23] is utilized, where the VLMs are supplied with the input query image and a manufacturing feature list, along with a single illustrative example. This example, shown in Fig. 7, includes a CAD design and a structured JSON output detailing the ground truth features with their names and quantities. This comprehensive example includes all potential manufacturing features to aid the VLMs in learning and recognizing these features in unseen CAD images.

**Few-shot Learning Prompt**

```
###Example Task:
Before proceeding with your task, here is an example to illustrate the expected format and
content of your response.

(Example image and answer will be provided below.)
```

**Example image**

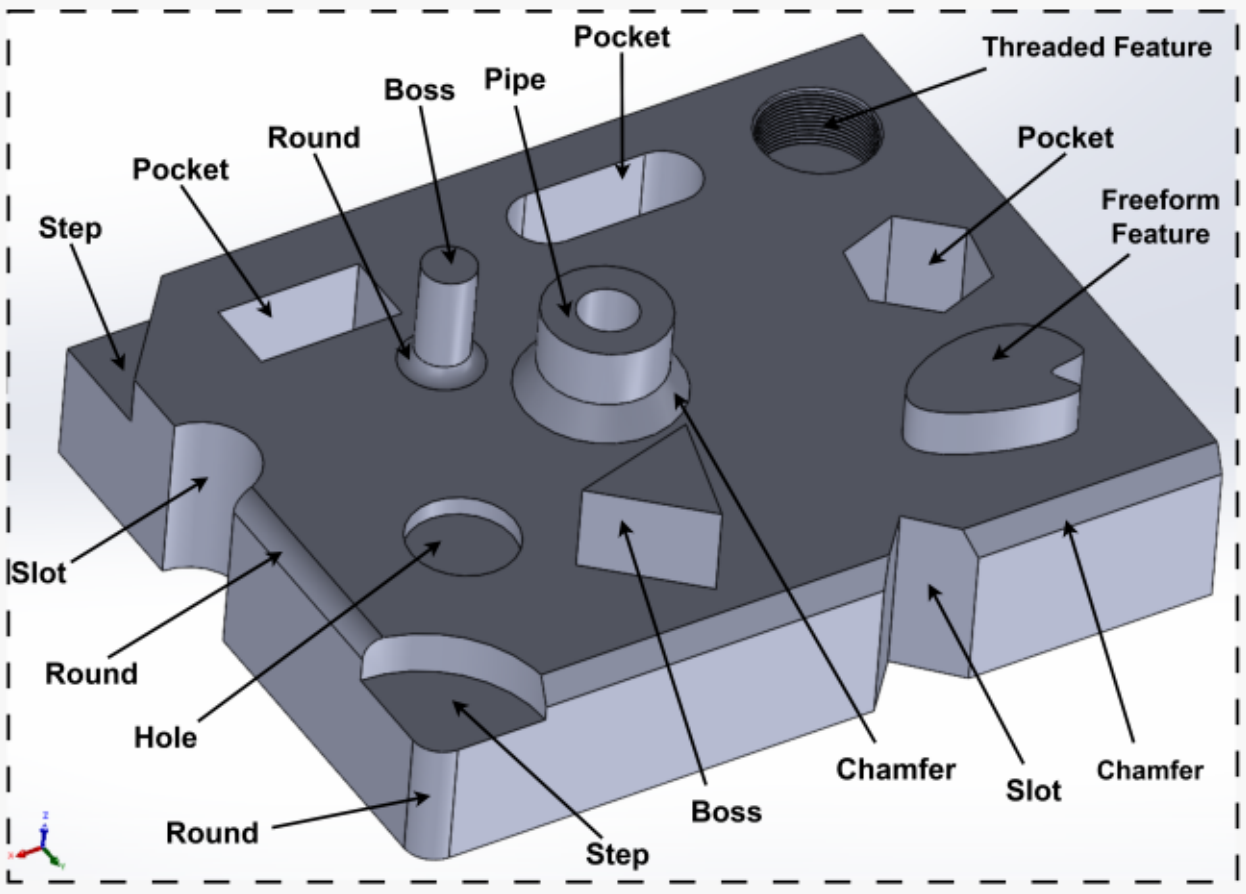

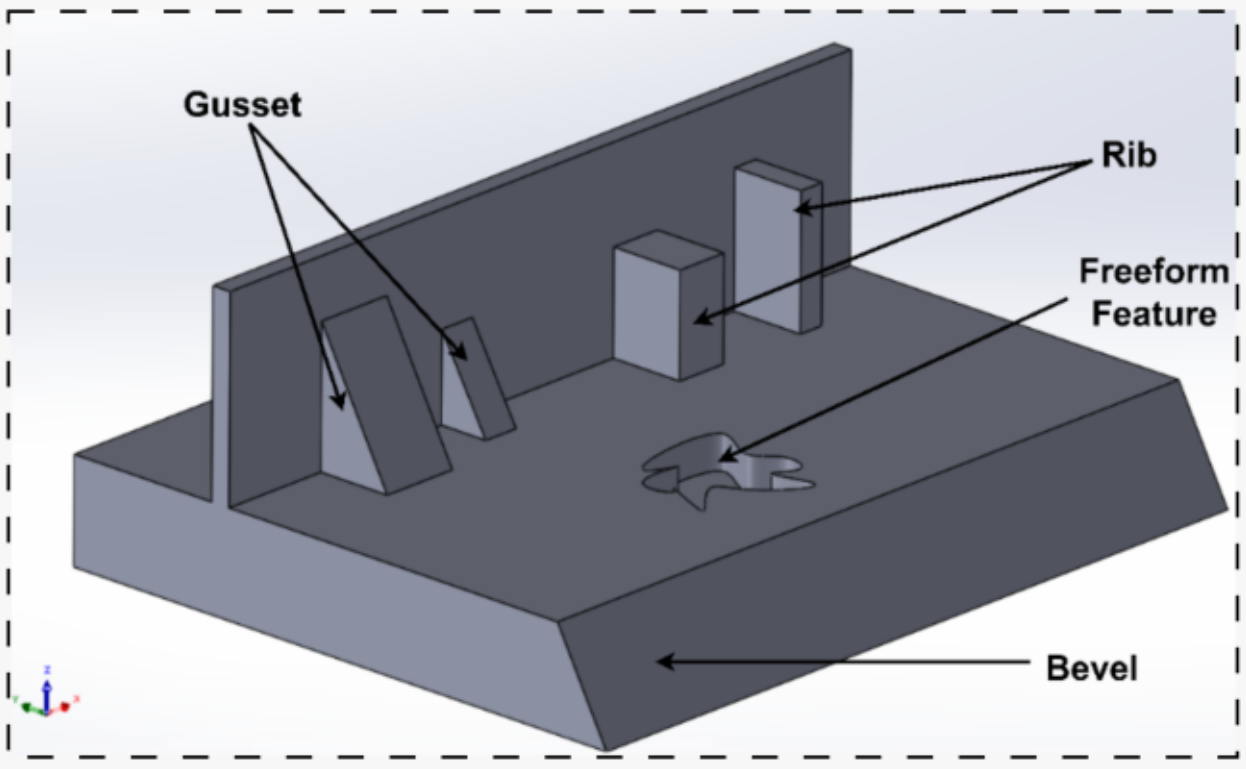

**Example answer**

```
{
  "identified_features": [ {
      "feature_name": "Hole (Through / Blind Hole)",
      "exists": true,
      "quantity": 2
    },
    {
      "feature_name": "Slot (Through / Blind)",
      "exists": true,
      "quantity": 2
    },
    {
      "feature_name": "Step (Through / Blind Step)",
      "exists": true,
      "quantity": 2
    },
    {
      "feature_name": "Pocket (Blind / Through)",
      "exists": true,
      "quantity": 3
    },
    {
      "feature_name": "Chamfer / Bevel",
      "exists": true,
      "quantity": 2
    },
    {
      "feature_name": "Fillet / Round",
      "exists": true,
      "quantity": 3
    },
    {
      "feature_name": "Threaded Feature",
      "exists": true,
      "quantity": 1
    },
    ...
  ]
}
```

**Fig. 7 Few-shot learning prompt.**

### 3.3.4. Parallel Prompt vs. Sequential Prompt

The structuring of input prompts, either parallel or sequential, serves as useful techniques to effectively query VLMs in identifying manufacturing features, as illustrated in Fig. 8.

Parallel prompting presents all relevant instructions to the VLMs simultaneously [93]. This approach allows the VLMs to access all necessary data at once, enhancing processing efficiency and reducing response times. As demonstrated in Fig. 8, the VLM receives a complete list of features along with instructions to identify feature names and quantities in CAD designs. Although this method reduces the number of prompts needed and improves scalability for managing multiple feature categories, it may also increase the likelihood of model errors or hallucinations due to the complexity of processing multiple inputs at the same time [64].

Sequential prompting requires VLMs to process information in stages, building upon each previous step [93]. This method is particularly effective for managing complex information. In this study, sequential prompting is applied in a structured manner: the VLM first reviews a comprehensive manufacturing features list, and then identifies the presence and quantity of each feature within the CAD images. This systematic assessment of each feature tends to enhance accuracy, though it may increase processing time.

**Parallel Prompt**

```
Return the results in JSON format.
**Do not include any other content, including any punctuation around the text.**
Example output format:
{{
    "identified_features": [
        {{
            "feature_name": "Hole (Through / Blind Hole)",
            "exists": true,
            "quantity": 2
        }},
        {{
            "feature_name": "Fillet / Round (Concave / Convex)",
            "exists": false,
            "quantity": 0
        }}
    ]
}}
```

**Sequential Prompt**

```
Instructions:
1. **Review the COMPLETE list of manufacturing features one by one.**
2. **Determine whether each feature exists in the given images.**
3. **Consider the **hierarchy** of each feature name from the provided list to understand its meaning in the manufacturing context.**
4. **Identify features using the provided list, adhering strictly to the naming conventions.**
5. **Examine each view of the part to identify all features. Different angles may reveal features not visible in other views.**
6. **For each identified feature, provide the quantity:** The number of such features present in the images (e.g., if there are two "Through Holes", then the quantity is 2). This should be a numerical integer (0, 1, 2, ...).
```

**Fig. 8 Parallel prompts vs. Sequential prompts.**

### 3.3.5. Chain-of-Thoughts (CoT)

Previous research has shown that introducing intermediate reasoning steps, known as a chain-of-thought (CoT), significantly enhances the performance of LLMs in various reasoning tasks [94–97]. This approach has evolved into the tree-of-thoughts [98,99], which incorporates multiple reasoning pathways to enhance language models' performance [100,101].

In this work, a three-step CoT reasoning process is implemented. Initially, the VLM assesses the presence of a feature in the CAD design based on visual observations. It then considers the manufacturing context and feature hierarchy to evaluate how these factors influence feature recognition. Finally, the VLM counts the occurrences of the feature, providing an exact numerical output. This structured reasoning process, illustrated in Fig. 9, mirrors human cognitive approaches to problem-solving and ensures a detailed contextual analysis.

**Chain-of-Thoughts (CoT) Prompt**

```
Instructions:
1. **Go through the COMPLETE list of manufacturing features one by one.**
2. **For each feature, follow these steps to determine its existence and quantity:**
     **Step 1:** Identify the feature present in the CAD image. Describe what you see that
makes you think the feature exists or does not exist.
     **Step 2:** Consider the manufacturing context and hierarchy. Think about how this
context influences your decision.
     **Step 3:** Determine the quantity of the feature. Count the occurrences of this feature
in the image.
3. **Provide your answers in a structured format for each feature, including your reasoning
for existence and quantity.**
4. **Ensure that each feature from the provided list has been considered in your final
output.**
5. **Keep the reasoning concise and to the point, no more than one or two sentences.**
```

**Fig. 9 Chain-of-thoughts prompt.**

## 3.4. Evaluation Metrics

To address the third research question regarding the effective quantification and evaluation of VLMs in AFR, this study employs four key evaluation metrics to accurately assess the performance of the VLMs.

### 3.4.1. Feature Name Accuracy (FNA)

Feature Name Accuracy (FNA) measures the correctness of the feature names identified by the VLM. It is defined as the ratio of correctly identified feature names to the total number of ground truth feature names in a CAD design:

$$FNA = \frac{Correctly\ Identified\ Names}{Total\ Ground\ Truth\ Names} \times 100\% \quad (1)$$

A high FNA value indicates that the identified features are correctly named. Accurate naming is crucial for ensuring that subsequent manufacturing processes are based on precise and reliable feature data.

### 3.4.2. Feature Quantity Accuracy (FQA)

Feature Quantity Accuracy (FQA) evaluates the VLM's ability to recognize the correct number of features. It is defined as the ratio of the true positive quantity of features to the ground truth quantity:

$$FQA = \frac{True\ Positive\ Quantity}{Ground\ Truth\ Quantity} \times 100\% \quad (2)$$

A high FQA value indicates that the model accurately identifies the correct number of each feature. This metric is crucial as it ensures accurate feature counts, optimizing manufacturing workflows and resource use, while reducing waste and enhancing production efficiency.

### 3.4.3. Hallucination Rate (HR)

Hallucination rate (HR) measures the frequency of incorrect or irrelevant features generated by the VLM. It is defined as the ratio of hallucinated features to the total predicted features.

$$HR = \frac{Hallucinated\ Quantity}{Predicted\ Quantity} \times 100\% \quad (3)$$

A low HR value indicates fewer incorrect predictions, critical for accurate cost estimations and efficient process planning, avoiding costly manufacturing errors.

### 3.4.4. Mean Absolute Error (MAE)

Mean absolute error (MAE) measures the difference between the predicted feature quantities and the actual quantities. It is calculated as the average of the absolute differences between the predicted and ground truth quantities for all features in a CAD design.

$$MAE = \frac{1}{n}\sum_{i=1}^{n} |Q_i - \widehat{Q_i}| \tag{4}$$

where $n$ is the total number of features being evaluated, $Q_i$ is the ground truth quantity of the $i$th feature, $\widehat{Q_i}$ is the predicted quantity of the $i$th feature, and $|Q_i - \widehat{Q_i}|$ represents the absolute error of the $i$th prediction. This metric provides a clear measure of prediction accuracy. A small MAE value, closer to 0, indicates high accuracy in predicting feature quantities.

The four metrics serve as benchmarks to compare the VLMs' performance against human experts in feature labelling accuracy, quantity estimation, and error minimization.

# 4. Results and Discussion

This section presents an in-depth evaluation of five VLMs across six distinct experiments, designed to test their capability in recognizing manufacturing features in CAD designs. The analysis spans three levels of CAD design complexity and employs a variety of prompt engineering techniques to assess VLMs' performance. Key evaluation metrics are used to compare the effectiveness of each VLM.

## 4.1. Overall VLMs' Performance Evaluation

The evaluation of five VLMs across six distinct experiments involving 100 CAD designs has yielded several key insights, as shown in Fig. 10. Claude-3.5 achieves the highest feature quantity accuracy of 74% using a zero-shot and multi-view prompt strategy (Experiment 3). Claude-3.0, while outperforming open-source VLMs with an accuracy of 40%, does not match the performance of Claude-3.5. Open-source models consistently show lower accuracies (<40%). Claude-3.5 shows strong performance with zero-shot and multi-view prompts, indicating that well-crafted zero-shot prompts can sometimes outperform few-shot prompts. This may be due to the fact that few-shot prompts can introduce ambiguity by being misinterpreted as part of a narrative rather than clear instructions [102].

Both GPT-4o and Claude-3.5 achieve a feature name accuracy of 75% in their optimal configurations: GPT-4o with a multi-view and Few-shot-CoT setup (Experiment 6), and Claude-3.5 with a multi-view and Zero-shot-CoT approach (Experiment 5). GPT-4o's success in few-shot learning underscores its ability to generalize effectively when provided with relevant labelled examples. This aligns with research indicating that few-shot prompts are most effective when the in-context examples closely align with the test cases [103–105].

In terms of hallucination rate, GPT-4o consistently leads across all experiments with the lowest rate of 8% in a few-shot with multi-view configuration (Experiment 4). Claude-3.5 shows slightly higher hallucinations at 12%, while open-source models exhibit the highest rates, exceeding 30%. This trend suggests a performance decline in open-source models as prompt complexity increases, likely due to their smaller context windows [106] compared to closed-source models.

Regarding mean absolute error (MAE), Claude-3.5 records the lowest value at 3.2 in a zero-shot with multi-view setup (Experiment 3), outperforming all other VLMs. Claude-3.0 attains a slightly better MAE value at 6.4 compared to open-source models but remains less effective than the other closed-source models. The open-source models show high MAE values as high as 12.0.

Figure 11 provides a comparative analysis of the top-performing experiments for each VLM. Overall, Claude-3.5 in the zero-shot with multi-view setup (Experiment 3) emerges as the superior model based on accuracy and MAE scores, while GPT-4o is particularly effective in minimizing hallucinations. These findings highlight the distinct strengths of Claude-3.5 and GPT-4o across different evaluation metrics.

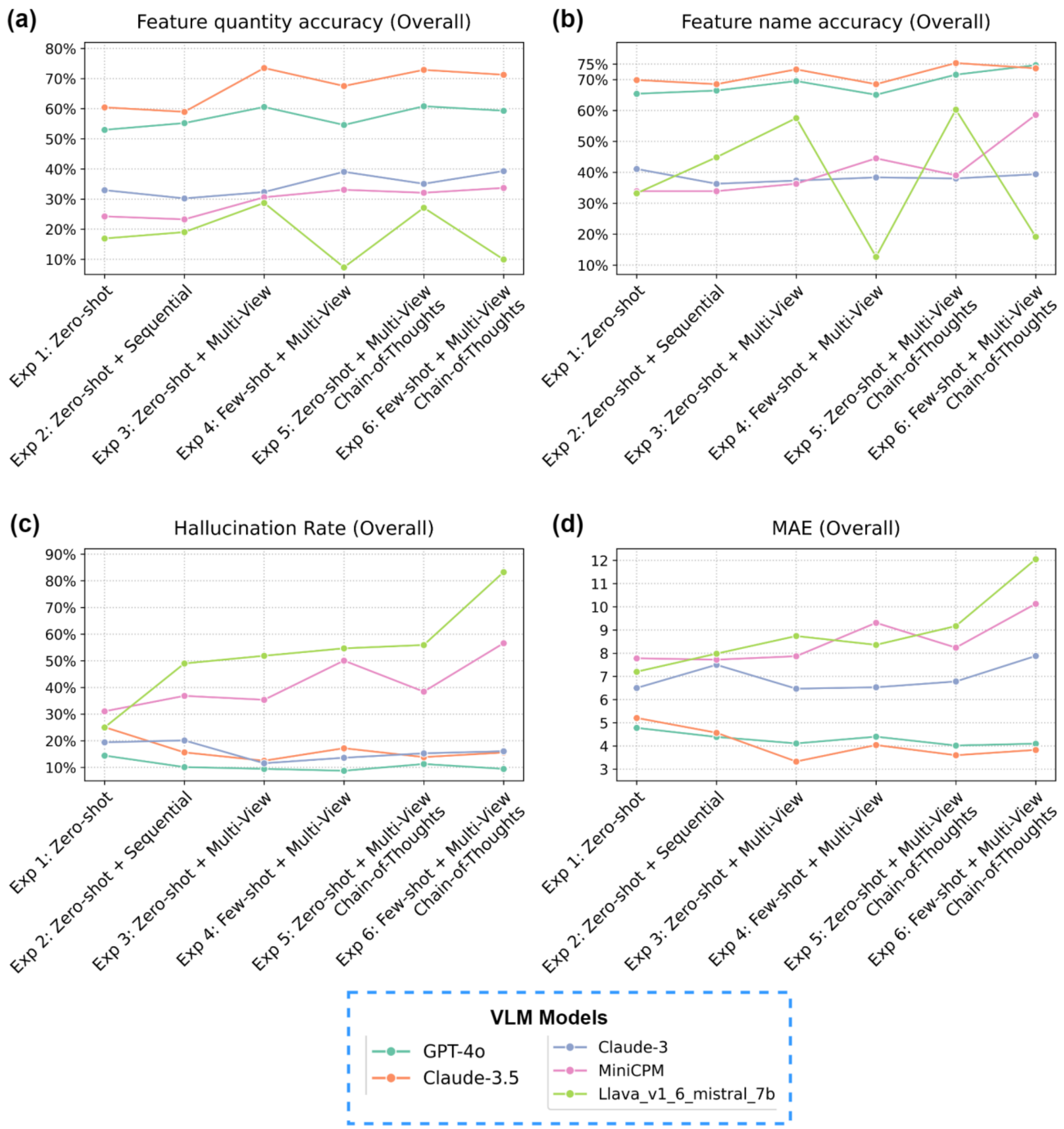

**Fig. 10 Overall performance of five VLMs across six experiments for 100 CAD models. Claude-3.5 outperforms all other VLMs in feature quantity accuracy, feature name accuracy, and MAE. Conversely, GPT-4o excels in minimizing hallucination rates, while open-source VLMs demonstrate the lowest overall performance.**

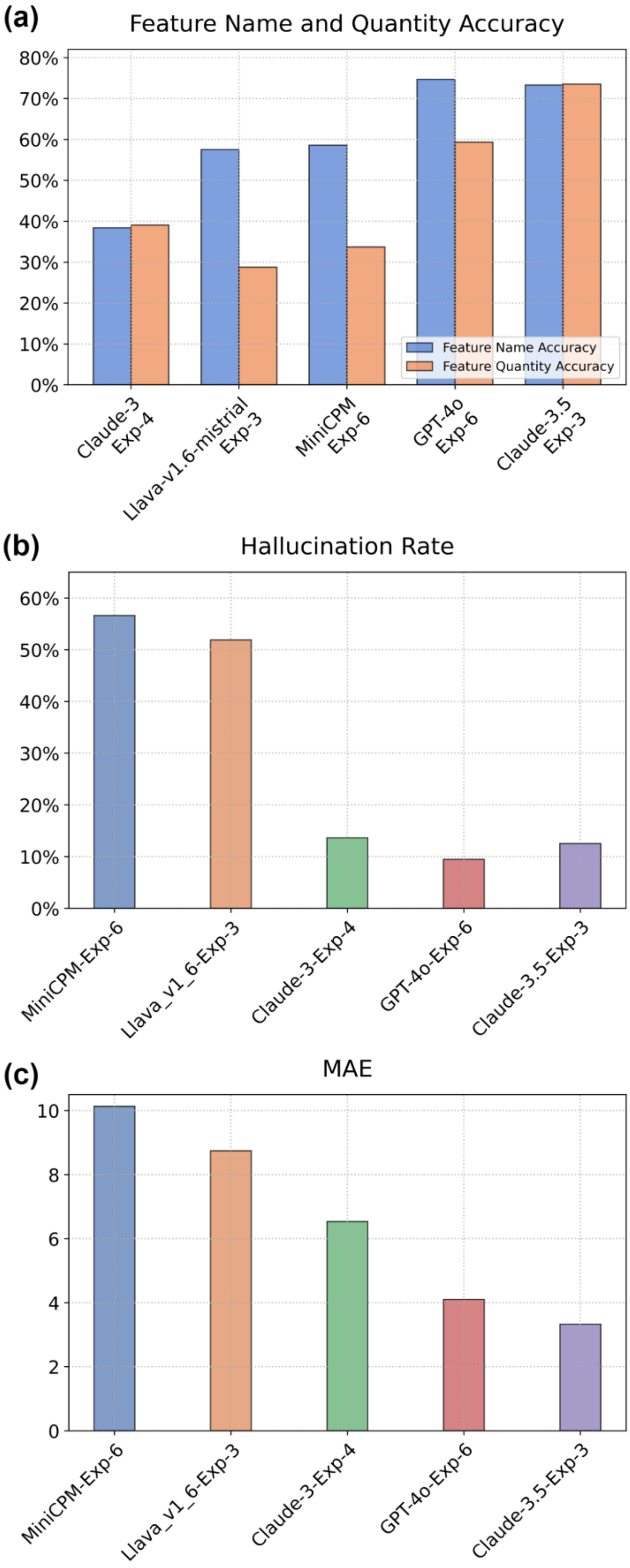

**Fig. 11 Performance comparison between the top performing experiments under each VLM. Claude-3.5 leads in feature quantity accuracy and MAE in Experiment 3, whereas GPT-4o excels in feature name accuracy and exhibits the least hallucinations.**

## 4.2. Performance Evaluation on Easy-difficulty CAD Designs

The performance analysis of five distinct VLMs on 33 easy-difficulty CAD designs is summarized in Fig. 12. GPT-4o achieves the highest feature quantity accuracy at 79% using the multi-view with Few-shot-CoT prompt (Experiment 6), while the open-source models demonstrate the lowest accuracy in this category. GPT-4o also leads in feature name accuracy, achieving 83% in both Experiments 4 and 6, likely benefiting from the effective use of closely matching example in few-shot learning.

In terms of hallucination rates, GPT-4o consistently records the lowest rate of 9% across multiple experiments. Although Claude-3.5 shows slightly higher hallucination rates than GPT-4o, both models significantly outperform the open-source models, which exhibit the highest rates. GPT-4o also records the lowest MAE of 0.65 in Experiment 5, maintaining strong performance across other setups, while the open-source models display the highest MAE values.

Figure 13 showcases examples of VLM-generated results for easy CAD designs. GPT-4o, particularly in Experiment 6, stands out as the top-performing model on this test set, excelling across all evaluation metrics. Notably, all VLMs frequently struggle to differentiate between '*chamfer*' and '*fillet*' features, likely due to their subtle visual differences. Moreover, open-source models tend to generate more hallucinated features compared to their closed-source counterparts. Unlike GPT-4o, which tends to underestimate features leading to conservative estimates, Claude-3.5 shows slightly higher hallucination rates, risking overestimations that could inflate costs in downstream tasks. Additional visual illustrations of feature predictions by VLMs on easy-level CAD designs can be found in Fig. B1 of Appendix B.

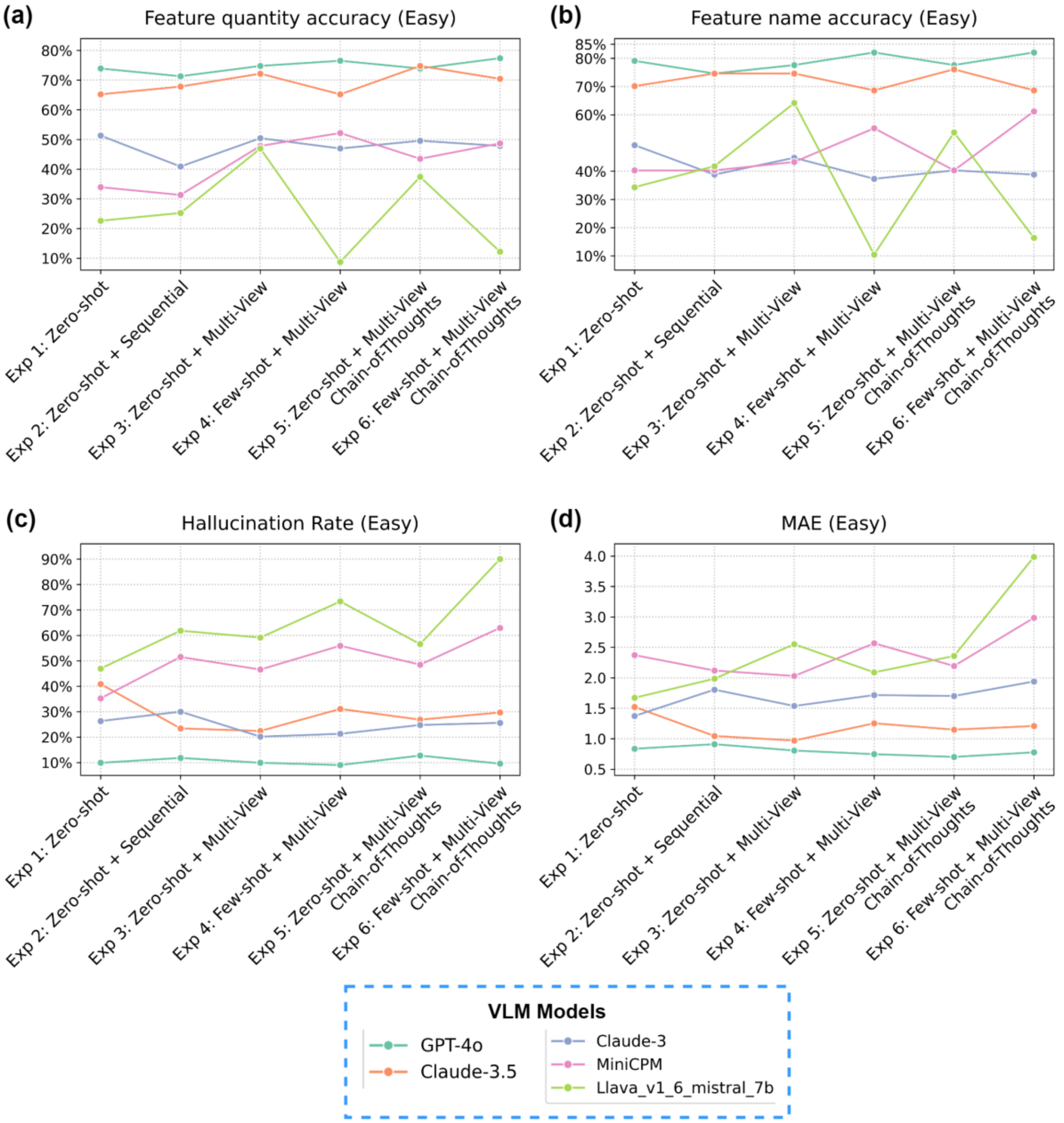

**Fig. 12 Performance evaluation of five VLMs across six experiments for 33 easy CAD models. GPT-4o, particularly in Experiment 6, outperforms all other VLMs across all evaluation metrics.**

**Manufacturing Feature Recognition on Easy-Level CAD Images**

| | Example 1 | Example 2 | Example 3 |
|---|---|---|---|
| Isometric View | 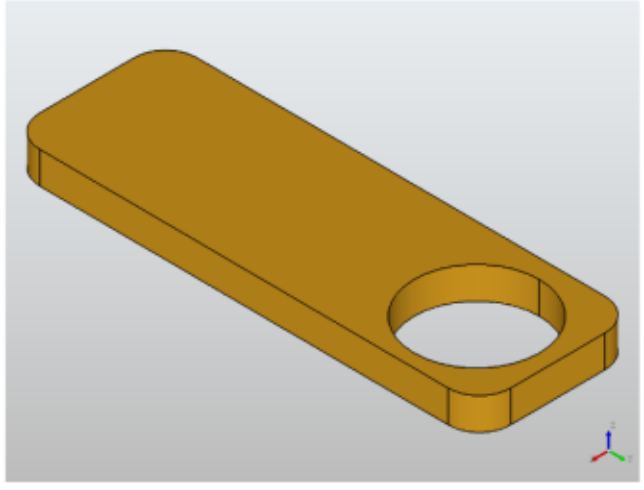 | 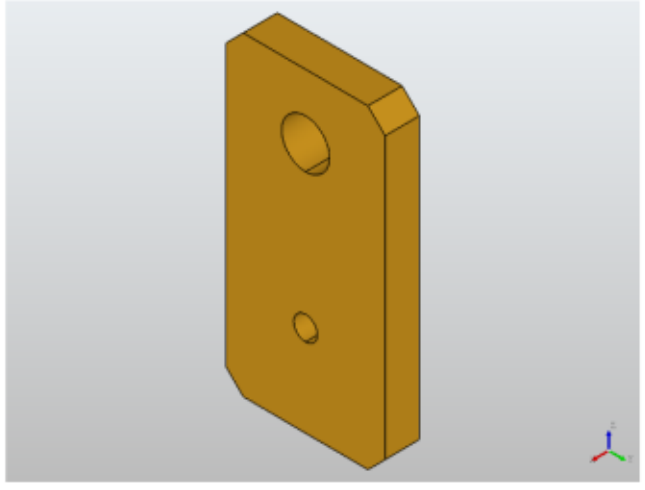 | 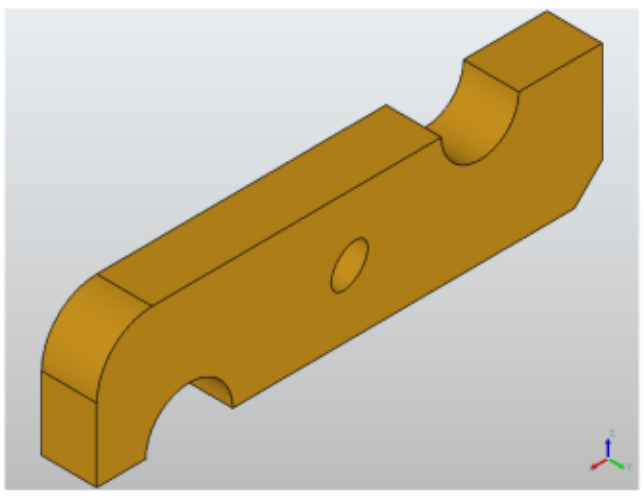 |
| Additional Views | 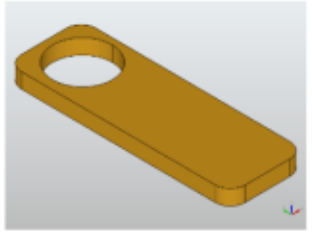 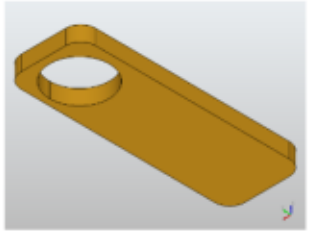 | 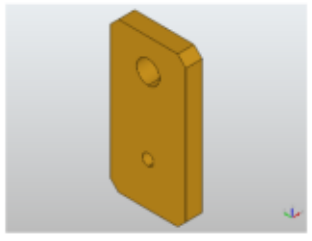 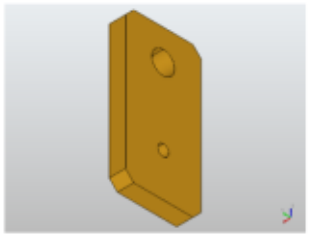 | 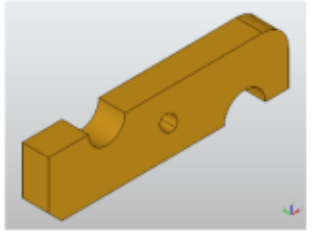 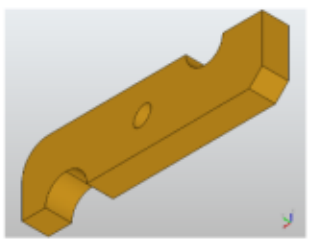 |
| **Ground Truth** (by human expert) | • Hole (through/blind): 1<br>• Fillet/Round: 4 | • Hole (through/blind): 2<br>• Chamfer/Bevel: 4 | • Hole (through/blind): 1<br>• Fillet/Round: 1<br>• Chamfer/Bevel: 1<br>• Slot (through/blind): 2 |
| **GPT-4o** Exp 6: Few-shot, Multi-view, CoT | • **Hole (through/blind): 1**<br>• **Fillet/Round: 4** | • **Hole (through/blind): 2**<br>• **Chamfer/Bevel: 2** | • **Hole (through/blind): 1**<br>• **Fillet/Round: 2**<br>• **Chamfer/Bevel: 0**<br>• **Slot (through/blind): 0** |
| **Claude-3.5** Exp 5: Zero-shot, Multi-view, CoT | • **Hole (through/blind): 1**<br>• **Fillet/Round: 8**<br>• **Chamfer/Bevel: 8** | • **Hole (through/blind): 2**<br>• **Chamfer/Bevel: 1**<br>• **Step (through/blind): 1** | • **Hole (through/blind): 1**<br>• **Fillet/Round: 2**<br>• **Slot (through/blind): 2**<br>• **Step (through/blind): 2**<br>• **Chamfer/Bevel: 0** |
| **Claude-3** Exp 3: Zero-shot, Multi-view | • **Hole (through/blind): 2**<br>• **Fillet/Round: 4** | • **Hole (through/blind): 2**<br>• **Chamfer/Bevel: 0** | • **Hole (through/blind): 2**<br>• **Fillet/Round: 1**<br>• **Chamfer/Bevel: 0**<br>• **Slot (through/blind): 0** |
| **MiniCPM-Llama3** Exp 4: Few-shot, Multi-view | • **Hole (through/blind): 2**<br>• **Fillet/Round: 0**<br>• **Chamfer/Bevel: 1** | • **Hole (through/blind): 3**<br>• **Chamfer/Bevel: 1** | • **Hole (through/blind): 2**<br>• **Slot (through/blind): 3**<br>• **Chamfer/Bevel: 1**<br>• **Fillet/Round: 0**<br>• **Step (through/blind): 1**<br>• **Pocket (through/blind): 1**<br>• **Boss: 1**<br>• **Freeform features: 1** |
| **Llava-1.6-mistral-7b** Exp 3: Zero-shot, Multi-view | • **Hole (through/blind): 2**<br>• **Fillet/Round: 0**<br>• **Chamfer/Bevel: 1** | • **Hole (through/blind): 2**<br>• **Chamfer/Bevel: 1**<br>• **Slot (through/blind): 1**<br>• **Step (through/blind): 1**<br>• **Pocket (through/blind): 1**<br>• **Neck: 1** | • **Hole (through/blind): 2**<br>• **Chamfer/Bevel: 1**<br>• **Fillet/Round: 0**<br>• **Slot (through/blind): 0** |

*Note: **green** colour represents correct prediction; **red** colour represents false prediction or misprediction; **orange** colour represents hallucination*

**Fig. 13 Examples of ground truth and VLM-generated features for easy-level CAD designs. GPT-4o in Experiment 6 shows the highest accuracy and the lowest rate of hallucinated features compared to other setups.**

## 4.3. Performance Evaluation on Medium-difficulty CAD Designs

The performance analysis of five VLMs on 33 medium-difficulty CAD designs is summarized in Fig. 14. Claude-3.5 achieves the highest feature quantity accuracy of 74.5% in Experiment 5. Both Claude-3.5 and GPT-4o excel in feature name accuracy, each securing a maximum accuracy of 74.9%. In contrast, Llava-v1.6-mistral-7b shows the lowest accuracies, as low as 10% in feature quantity and 15% in feature name accuracy.

Regarding hallucination rates, GPT-4o demonstrates the lowest rate at 8% across several experiments. The open-source models, however, exhibit higher hallucination rates (>30%), indicating difficulties with the increased complexity of the medium-level CAD designs. Furthermore, GPT-4o records the lowest MAE at 0.7 in Experiment 6, maintaining strong performance across other setups, whereas the open-source models display significantly higher MAE values, exceeding 2.0.

Figure 15 presents visual examples of predictions on medium-level CAD designs, highlighting GPT-4o in Experiment 6 as the top performer in minimizing errors and hallucinations. Notably, all VLMs struggle to differentiate between '*pipe/tube*' and '*boss*' features likely due to their visual similarities. The open-source models particularly perform poorly, often misidentifying or hallucinating features across the CAD designs. Additional visual examples of feature predictions by VLMs on medium-level CAD designs can be found in Fig. B2 of Appendix B.

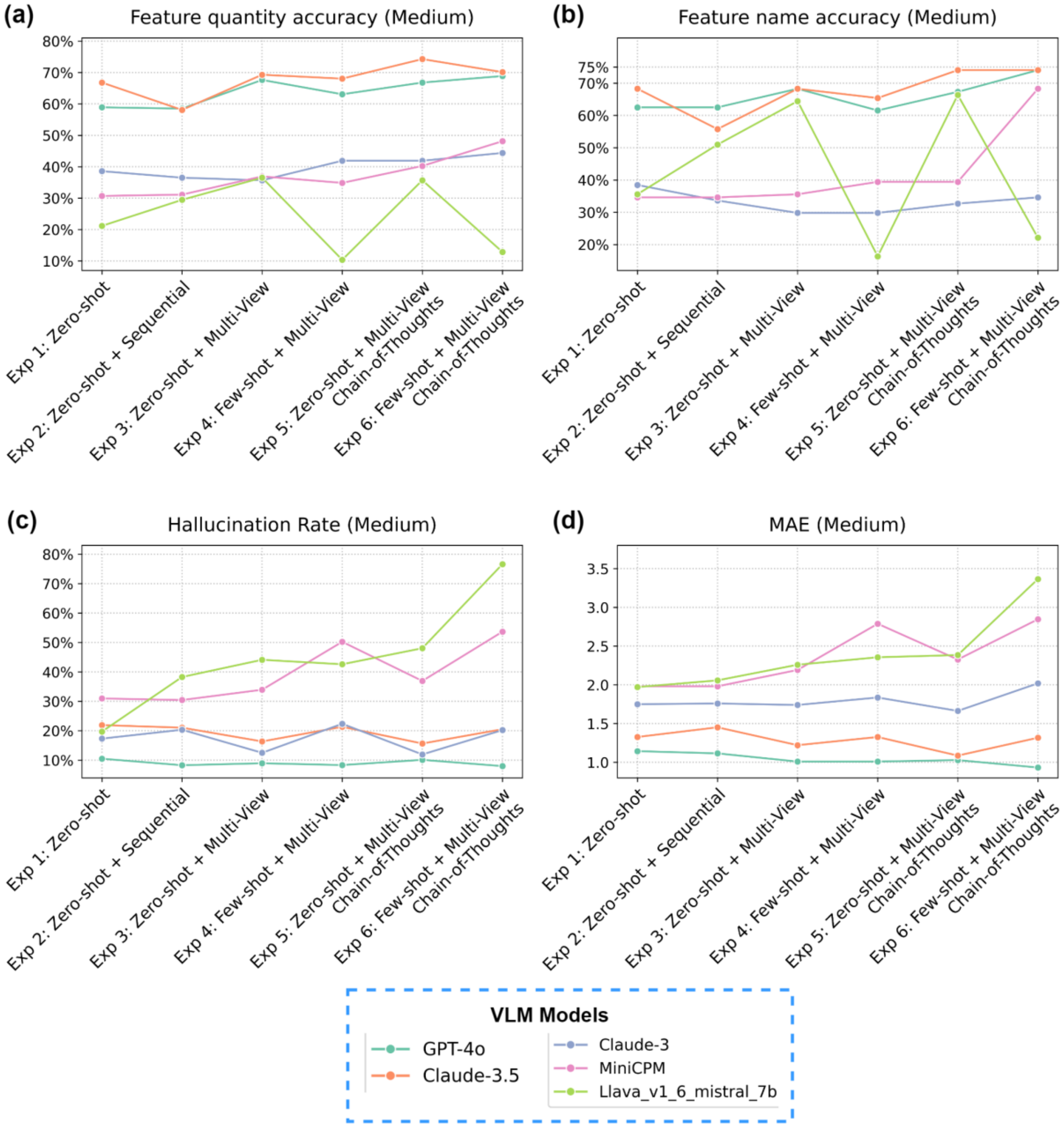

**Fig. 14 Performance evaluation of five VLMs across six experiments for 33 medium CAD models. Claude-3.5 and GPT-4o show high accuracy in feature name and quantity, significantly outperforming other VLMs, and exhibit the lowest rates of hallucinated features.**

| Manufacturing Feature Recognition on Medium-Level CAD Images | | | |
|---|---|---|---|
| | Example 1 | Example 2 | Example 3 |
| Isometric View | 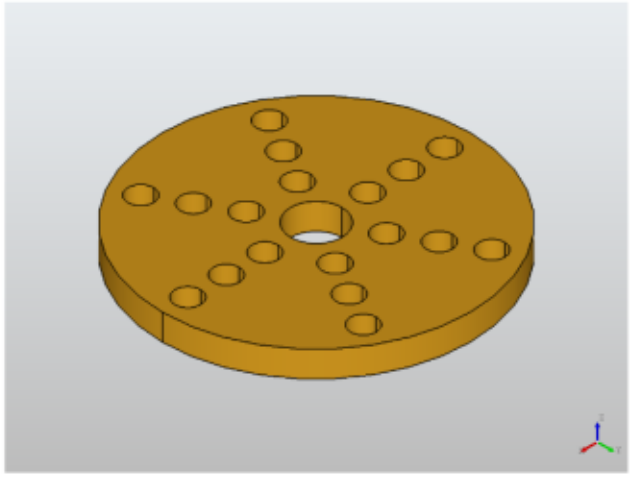 | 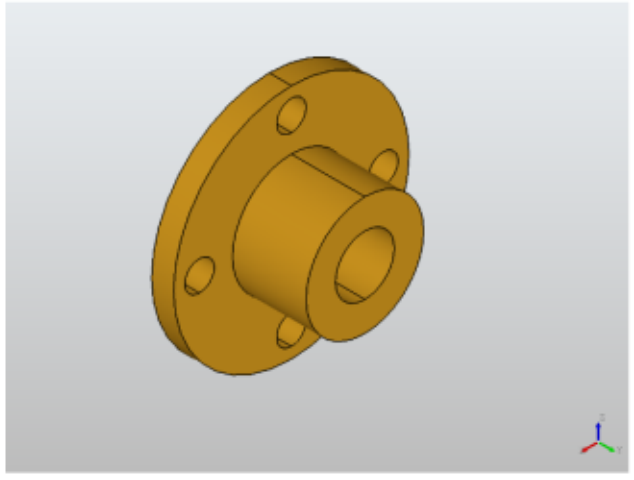 | 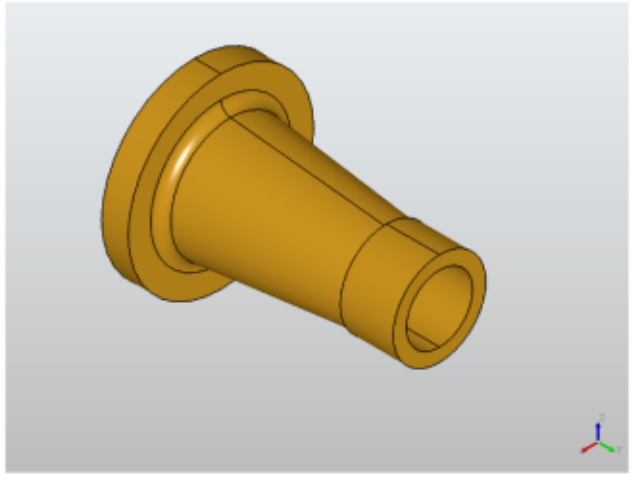 |
| Additional Views | 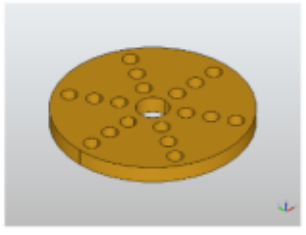 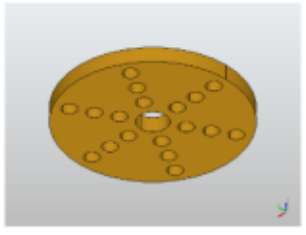 | 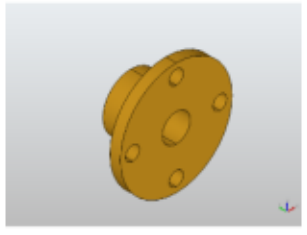 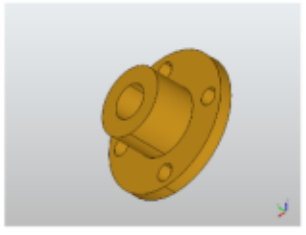 | 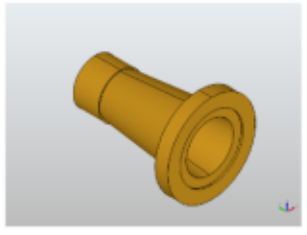 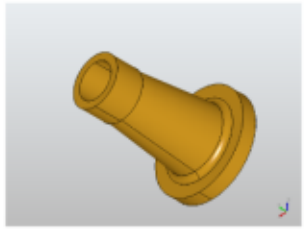 |
| **Ground Truth** (by human expert) | • Hole (through/blind): 19 | • Hole (through/blind): 5<br>• Pipe/Tube: 1 | • Hole (through/blind):1<br>• Boss: 1<br>• Fillet/Round: 1<br>• Pipe/Tube: 1 |
| **GPT-4o** Exp 6: Few-shot, Multi-view, CoT | • **Hole (through/blind): 18** | • **Hole (through/blind): 5**<br>• **Boss: 1**<br>• **Pipe/Tube: 0** | • **Hole (through/blind):1**<br>• **Boss: 1**<br>• **Fillet/Round: 1**<br>• **Pipe/Tube: 0** |
| **Claude-3.5** Exp 3: Zero-shot, Multi-view, CoT | • **Hole (through/blind): 20** | • **Hole (through/blind): 5**<br>• **Boss: 1**<br>• **Pipe/Tube: 0** | • **Hole (through/blind):1**<br>• **Boss: 1**<br>• **Fillet/Round: 2**<br>• **Pipe/Tube: 0** |
| **Claude-3** Exp 5: Zero-shot, Multi-view, CoT | • **Hole (through/blind): 19**<br>• **Fillet/Round: 1** | • **Hole (through/blind): 6**<br>• **Pipe/Tube: 0** | • **Fillet/Round: 2**<br>• **Hole (through/blind):0**<br>• **Boss: 0**<br>• **Pipe/Tube: 0** |
| **MiniCPM-Llama3** Exp 6: Few-shot, Multi-view, CoT | • **Hole (through/blind): 10**<br>• **Pocket (through/blind): 1**<br>• **Chamfer/Bevel: 2** | • **Hole (through/blind): 4**<br>• **Pocket (through/blind): 1**<br>• **Chamfer/Bevel: 2**<br>• **Boss: 2**<br>• **Pipe/Tube: 0** | • **Hole (through/blind): 1**<br>• **Boss: 1**<br>• **Pocket (through/blind): 1**<br>• **Chamfer/Bevel: 2**<br>• **Freeform Features: 1**<br>• **Step (through/blind): 2**<br>• **Pipe/Tube: 0**<br>• **Fillet/Round: 0** |
| **Llava-1.6-mistral-7b** Exp 5: Zero-shot, Multi-view, CoT | • **Hole (through/blind): 2**<br>• **Pocket (through/blind): 1**<br>• **Fillet/Round: 1** | • **Hole (through/blind): 2**<br>• **Fillet/Round: 1**<br>• **Pipe/Tube: 0** | • **Hole (through/blind): 2**<br>• **Fillet/Round: 1**<br>• **Chamfer/Bevel: 1**<br>• **Threaded Features: 1**<br>• **Neck: 1**<br>• **Pipe/Tube: 0**<br>• **Boss: 0** |

*Note:* ***green*** *colour represents correct prediction;* ***red*** *colour represents false prediction or misprediction;* ***orange*** *colour represents hallucination*

**Fig. 15 Examples of ground truth and VLM-generated answers for medium-level CAD designs. GPT-4o and Claude-3.5 both show high accuracy, and the lowest rate of hallucinated features compared to other VLMs.**

## 4.4. Performance Evaluation on Challenging CAD Designs

The evaluation of five VLMs on 34 challenging CAD designs is detailed in Fig. 16. Claude-3.5 stands out by achieving the highest feature quantity accuracy at 76% and feature name accuracy at 77% in Experiment 3. In contrast, Llava-v1.6-mistral-7b records the lowest accuracies, with only 6% in feature quantity and 10% in feature name accuracy.

Claude-3 achieves the lowest hallucination rate of 9% in Experiment 4, closely followed by Claude-3.5. Additionally, Claude-3.5 shows the lowest MAE, scoring 1.4 in Experiment 3, whereas the open-source VLMs exhibit considerably higher hallucination rates (>28%) and the highest MAE values (>3.3). This highlights a significant performance gap between open-source and closed-source models in analyzing challenging CAD designs.

Figure 17 presents visual examples of predictions on these designs, demonstrating that Claude-3.5 in Experiment 3 excels across all metrics. This VLM's proficiency is especially apparent in its ability to closely match the ground truth in designs with high feature counts, such as predicting 114 out of 119 '*blind holes*' in one example. The overall high accuracy in recognizing challenging parts suggests that tasks difficult for human experts might be more manageable for VLMs. Additional visual examples of VLMs predictions on challenging CAD designs can be found in Fig. B3 of Appendix B.

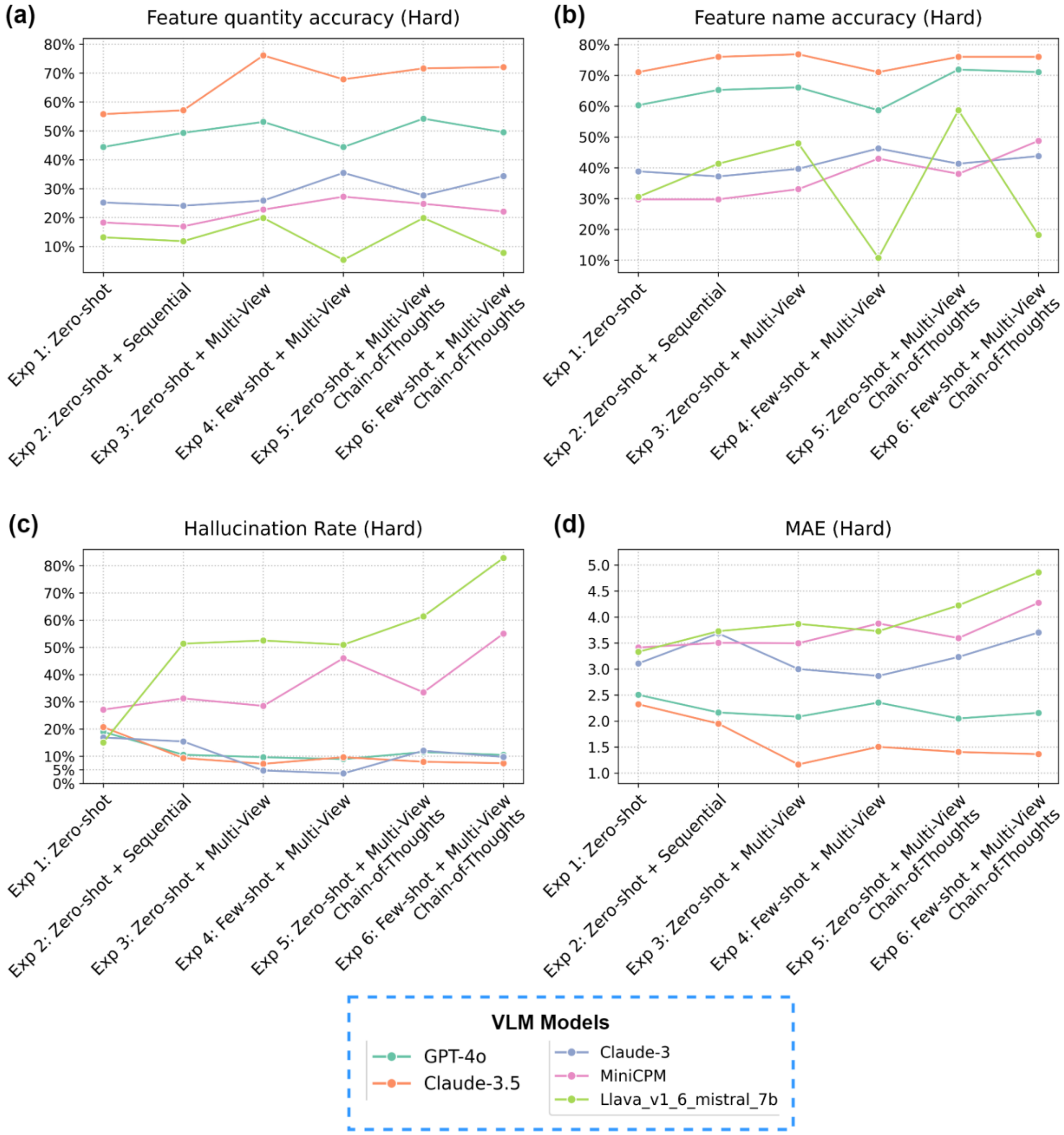

**Fig. 16 Performance evaluation of five VLMs across six experiments for 34 challenging CAD models. Claude-3.5 in Experiment 3 demonstrates high accuracy in feature name and quantity and exhibits the lowest rates of hallucinated features.**

| Manufacturing Feature Recognition on Difficult-Level CAD Images | | | |
|---|---|---|---|
| | Example 1 | Example 2 | Example 3 |
| Isometric View | 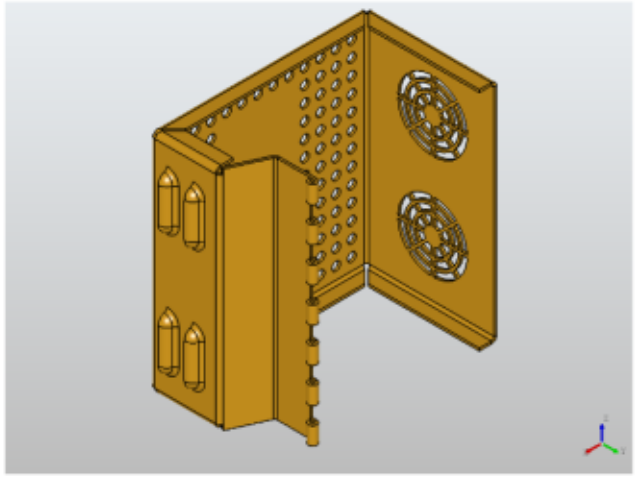 | 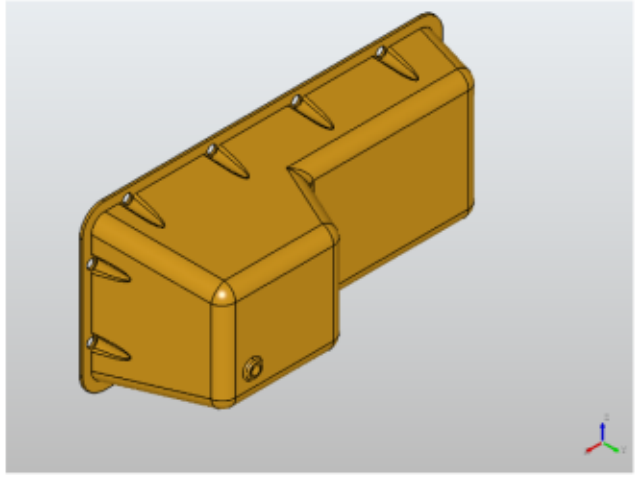 | 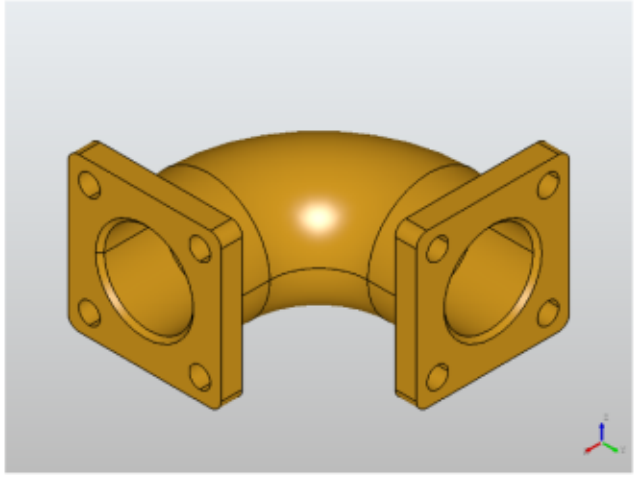 |
| Additional Views | 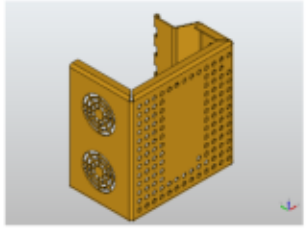 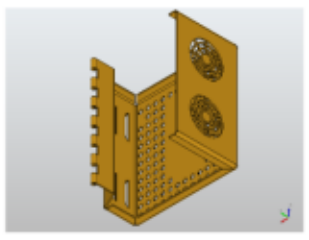 | 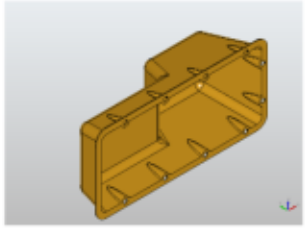 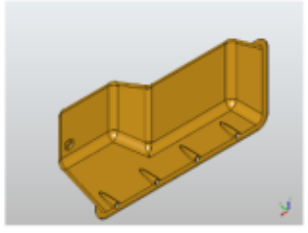 | 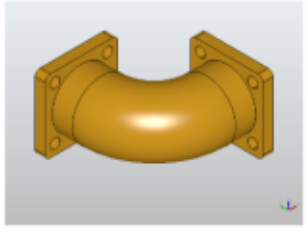 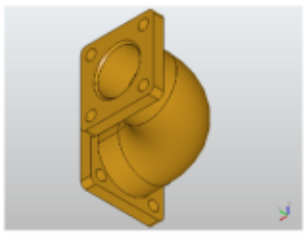 |
| **Ground Truth** (by human expert) | • Hole (through/blind): 119<br>• Slot (through/blind): 4<br>• Step (through/blind): 1<br>• Sheet metal: 1 | • Hole (through/blind): 11<br>• Step (through/blind): 1<br>• Fillet/Round: 4<br>• Sheet metal: 1 | • Hole (through/blind): 8<br>• Sheet metal: 1<br>• Pipe/Tube: 1<br>• Fillet/Round: 2 |
| **Claude-3.5** Exp 3: Zero-shot, Multi-view | • **Hole (through/blind): 114**<br>• **Slot (through/blind): 4**<br>• **Step (through/blind): 1**<br>• **Sheet metal: 1** | • **Hole (through/blind): 10**<br>• **Step (through/blind): 1**<br>• **Fillet/Round: 4**<br>• **Sheet metal: 0** | • **Hole (through/blind): 8**<br>• **Fillet/Round: 2**<br>• **Pipe/Tube: 1**<br>• **Sheet metal: 0**<br>• **Boss: 2** |
| **GPT-4o** Exp 5: Zero-shot, Multi-view, CoT | • **Hole (through/blind): 2**<br>• **Slot (through/blind): 2**<br>• **Sheet metal: 1**<br>• **Freeform features: 4**<br>• **Step (through/blind): 0** | • **Hole (through/blind): 10**<br>• **Fillet/Round: 8**<br>• **Rib: 8**<br>• **Step (through/blind): 0**<br>• **Sheet metal: 0** | • **Hole (through/blind): 8**<br>• **Fillet/Round: 2**<br>• **Pipe/Tube: 1**<br>• **Sheet metal: 0** |
| **Claude-3** Exp 4: Few-shot, Multi-view | • **Hole (through/blind): 4**<br>• **Slot (through/blind): 0**<br>• **Step (through/blind): 0**<br>• **Sheet metal: 0** | • **Hole (through/blind): 4**<br>• **Fillet/Round: 8**<br>• **Step (through/blind): 0**<br>• **Sheet metal: 0** | • **Hole (through/blind): 2**<br>• **Fillet/Round: 4**<br>• **Pipe/Tube: 0**<br>• **Sheet metal: 0** |
| **MiniCPM-Llama3** Exp 5: Zero-shot, Multi-view, CoT | • **Hole (through/blind): 4**<br>• **Sheet metal: 1**<br>• **Chamfer/Bevel: 1**<br>• **Slot (through/blind): 0**<br>• **Step (through/blind): 0** | • **Hole (through/blind): 4**<br>• **Step (through/blind): 1**<br>• **Chamfer/Bevel: 2**<br>• **Fillet/Round: 0**<br>• **Sheet metal: 0** | • **Hole (through/blind): 4**<br>• **Fillet/Round: 1**<br>• **Step (through/blind): 1**<br>• **Chamfer/Bevel: 2**<br>• **Freeform Features: 1**<br>• **Sheet metal: 0**<br>• **Pipe/Tube: 0** |
| **Llava-1.6-mistral-7b** Exp 5: Zero-shot, Multi-view, CoT | • **Hole (through/blind): 2**<br>• **Slot (through/blind): 1**<br>• **Step (through/blind): 1**<br>• **Pocket (through/blind): 1**<br>• **Chamfer/Bevel: 1**<br>• **Fillet/Round: 1**<br>• **Sheet metal: 0** | • **Fillet/Round: 1**<br>• **Slot (through/blind): 1**<br>• **Pocket (through/blind): 1**<br>• **Chamfer/Bevel: 1**<br>• **Hole (through/blind): 0**<br>• **Step (through/blind): 0**<br>• **Sheet metal: 0** | • **Hole (through/blind): 1**<br>• **Fillet/Round: 1**<br>• **Boss: 1**<br>• **Sheet metal: 0**<br>• **Pipe/Tube: 0** |

*Note:* ***green*** *colour represents correct prediction;* ***red*** *colour represents false prediction or misprediction;* ***orange*** *colour represents hallucination*

**Fig. 17 Examples of ground truth and VLM-generated answers for difficult-level CAD designs. Claude-3.5 shows exceptional accuracy, closely matching high ground truth counts in certain designs.**

## 5. Conclusions

This research demonstrated the potential of VLMs for automatically recognizing diverse manufacturing features in CAD designs. By applying systematic prompt engineering techniques, the study enhanced VLMs' ability to interpret complex visual and textual data across various CAD design complexities. Comparative analysis revealed that Claude-3.5-Sonnet achieved the highest feature quantity accuracy (74%) and feature name matching accuracy (75%), while also maintaining the lowest MAE score (3.2). In contrast, GPT-4o exhibited the lowest hallucination rate (8%). The open-source models, Llava-v1.6-mistral-7b and MiniCPM-Llama3-V2.5, while accessible, encountered significant challenges with higher hallucination rates (>30%) and lower accuracies (<40%).

The main contributions of this study included the creation of a customized CAD dataset showcasing varying levels of feature complexity, the systematic implementation of prompt engineering strategies, and the proposal of tailored evaluation metrics for AFR tasks. The findings illustrated the potential of these VLMs to handle complex geometrical data that often challenge human experts, thus offering a viable path towards automation in manufacturing design.

The study showcased the application of VLMs in manufacturing feature recognition. However, there were several limitations that will be addressed in future research. The CAD designs, while categorized into three difficulty levels, did not fully represent the diversity and complexity of real manufacturing scenarios. They focused mainly on isolated feature recognition and lack the complexity of intersecting features or intricate assembly contexts. Future studies will incorporate a broader range of CAD designs to uncover potential biases. Another limitation was the VLMs' inability to extract geometric dimensions from recognized features, which are vital for downstream manufacturing tasks such as process selection, cost estimation, and quality control. Future research will therefore aim to integrate feature recognition with the interpretation of 2D engineering drawings to more effectively utilize geometric dimensions.

To fully leverage the capabilities of VLMs in real-world manufacturing, future work will focus on expanding the diversity of the CAD dataset, enhancing the VLMs' ability to extract and utilize geometric dimensions from 2D engineering drawings, and refining prompt engineering techniques. By addressing these areas, VLMs can be further improved, broadening their application in automated manufacturing processes and paving the way for more comprehensive manufacturing solutions.

## Declaration of Competing Interest

The authors declare that they have no known competing financial interests or personal relationships that could have appeared to influence the work reported in this paper.

## Acknowledgement

This work is supported by the Agency for Science, Technology and Research (A*STAR), Singapore through the RIE2025 MTC IAF-PP grant (Grant No. M22K5a0045). It is also supported by Singapore International Graduate Award (SINGA) (Awardee: Muhammad Tayyab Khan) funded by A*STAR and Nanyang Technological University, Singapore.

# Appendix

The appendix provides supplementary information supporting the findings presented in this study. It includes additional figures illustrating the performance of Vision-Language Models (VLMs) across various experimental setups based on prompt engineering techniques for manufacturing feature recognition in CAD designs.

## A. Prompt Engineering Experiments

**Prompt 1 (Baseline Prompt): Zero-shot, Single Query Image, Parallel Prompt**

```
You are provided with a CAD image. Your task is to identify and describe the manufacturing features present in this image. Use the provided list of feature names, which is delimited by triple backticks. Use this list as a reference and strictly follow the naming conventions:

'''{manufacturing_features_names}'''

Return the results in JSON format.
**Do not include any other content, including any punctuation around the text.**
Example output format:
{{
    "identified_features": [
        {{
            "feature_name": "Hole (Through / Blind Hole)",
            "exists": true,
            "quantity": 2
        }},
        {{
            "feature_name": "Fillet / Round (Concave / Convex)",
            "exists": false,
            "quantity": 0
        }}
    ]
}}
```

**Fig. A1 Zero-shot with a single view and parallel prompting strategy.**

**Prompt 2: Zero-shot, Single Query Image, Sequential Prompt**

```
You are provided with a CAD image. Your task is to identify and describe the manufacturing
features present in this image. Use the provided list of feature names, which is delimited by
triple backticks. Use this list as a reference and strictly follow the naming conventions:

'''{manufacturing_features_names}'''

Instructions:
1. **Go through the COMPLETE list of manufacturing features one by one.**
2. **Consider whether each particular feature exists in the image.**
3. **Consider the **hierarchy** of each feature name from the provided list to understand its
meaning in the manufacturing context.**
4. **Answer the features based on the provided list, with the naming conventions strictly
following the given list.**
5. For each feature, provide **Quantity:** The number of such features present in the image
(e.g., if there are two "Through Holes", then quantity is 2). This should be a numerical
integer (0, 1, 2, ...).

Return the results in the following JSON format, strictly following the specified
structure. **Do not include any other content, including any punctuation around the text.**:

Example output format:
{{
    "identified_features": [
        {{
            "feature_name": "Hole (Through / Blind Hole)",
            "exists": true,
            "quantity": 2
        }},
        {{
            "feature_name": "Fillet / Round (Concave / Convex)",
            "exists": false,
            "quantity": 0
        }}
    ]
}}
```

**Fig. A2 Zero-shot with a single view and sequential prompting technique.**

**Prompt 3: Zero-shot, Multi-view Query Images, Sequential Prompt**

```
You are provided with an image containing multiple views of a 3D CAD model taken from
different viewing angles. Your task is to identify and describe the manufacturing features
present in this image. Use the provided list of feature names, which is delimited by triple
backticks. Use this list as a reference and strictly follow the naming conventions:

'''{manufacturing_features_names}'''

Instructions:
1. **Go through the COMPLETE list of manufacturing features one by one.**
2. **Think about whether each particular feature exists in the given image.**
3. **When considering the existence of a feature, take the **hierarchy** of feature name from
the provided list into account. This will help you better understand the feature's  meaning in
the manufacturing context.**
4. **Answer the features based on the provided list, with the naming conventions strictly
following the given list.**
5. Examine **each views** of the part to identify all features. Different angles may reveal
features not visible in other views.
5. For each feature, provide **Quantity:** The number of such features present in the image
(e.g., if there are two "Through Holes", then quantity is 2). This should be a numerical
integer (0, 1, 2, ...).

Return the results in the following JSON format, strictly adhering to the specified
structure. **Do not include any other content, including any punctuation around the text.**:

Example output format:
{{
    "identified_features": [
        {{
            "feature_name": "Hole (Through / Blind Hole)",
            "exists": true,
            "quantity": 2
        }},
        {{
            "feature_name": "Fillet / Round (Concave / Convex)",
            "exists": false,
            "quantity": 0
        }}
    ]
}}
```

**Fig. A3 Zero-shot with multi-view and sequential prompting setup.**

**Prompt 4: Few-shot learning, Multi-view Query Images, Sequential Prompt**

```
You are provided with images containing multiple views of a 3D CAD model taken from different angles. Your task is to identify and describe the manufacturing features present in these images. Use the provided list of feature names, which is delimited by triple backticks. Use this list as a reference and strictly follow the naming conventions:

'''{manufacturing_features_names}'''

Instructions:
1. **Review the COMPLETE list of manufacturing features one by one.**
2. **Determine whether each feature exists in the given images.**
3. **Consider the **hierarchy** of each feature name from the provided list to understand its meaning in the manufacturing context.**
4. **Identify features using the provided list, adhering strictly to the naming conventions.**
5. **Examine each view of the part to identify all features. Different angles may reveal features not visible in other views.**
6. **For each identified feature, provide the quantity:** The number of such features present in the images (e.g., if there are two "Through Holes", then the quantity is 2). This should be a numerical integer (0, 1, 2, ...).

###Example Task:
Before proceeding with your task, here is an example to illustrate the expected format and content of your response.

(Example image and answer will be provided below.)
```

**Fig. A4 Few-shot with multi-view and sequential prompting setup.**

**Prompt 5: Zero-shot, Multi-view Query Images, Chain-of-Thoughts (CoT)**

```
You are provided with an image containing multiple views of a 3D CAD model taken from
different viewing angles. Your task is to identify and describe the manufacturing features
present in this image. Use the provided list of feature names, which is delimited by triple
backticks. Use this list as a reference and strictly follow the naming conventions:

'''{manufacturing_features_names}'''

Instructions:
1. **Go through the COMPLETE list of manufacturing features one by one.**
2. **For each feature, follow these steps to determine its existence and quantity:**
     **Step 1:** Identify the feature present in the CAD image. Describe what you see that
     makes you think the feature exists or does not exist.
     **Step 2:** Consider the manufacturing context and hierarchy. Think about how this
     context influences your decision.
     **Step 3:** Determine the quantity of the feature. Count the occurrences of this feature
     in the image. (e.g., if there are two "Through Holes", then the quantity is 2). This
     should be a numerical integer (0, 1, 2, ...).
3. **Ensure that each feature from the provided list has been considered in your final
output.**
4. **Keep the reasoning concise and to the point, no more than one or two sentences.**

Return the results in the following JSON format, strictly adhering to the specified
structure. **Do not include any other content, including any punctuation around the text.**:

Example output format:
{{
    "identified_features": [
        {{
            "feature_name": "Hole (Through / Blind Hole)",
            "exists": true,
            "quantity": 2
        }},
        {{
            "feature_name": "Fillet / Round (Concave / Convex)",
            "exists": false,
            "quantity": 0
        }}
    ]
}}
```

**Fig. A5 Zero-shot with multi-view and chain-of-thoughts (CoT) reasoning.**

**Prompt 6: Few-shot learning, Multi-view Query Images, Chain-of-Thoughts (CoT)**

```
You are provided with images containing multiple views of a 3D CAD model taken from different
angles. Your task is to identify and describe the manufacturing features present in these
images. Use the provided list of feature names, which is delimited by triple backticks. Use
this list as a reference and strictly follow the naming conventions:

'''{manufacturing_features_names}'''

Instructions:
1. **Go through the COMPLETE list of manufacturing features one by one.**
2. **For each feature, follow these steps to determine its existence and quantity:**
     **Step 1:** Identify the feature present in the CAD image. Describe what you see that
makes you think the feature exists or does not exist.
     **Step 2:** Consider the manufacturing context and hierarchy. Think about how this
context influences your decision.
     **Step 3:** Determine the quantity of the feature. Count the occurrences of this feature
in the image.
3. **Provide your answers in a structured format for each feature, including your reasoning
for existence and quantity.**
4. **Ensure that each feature from the provided list has been considered in your final
output.**
5. **Keep the reasoning concise and to the point, no more than one or two sentences.**

###Example Task:
Before proceeding with your task, here is an example to illustrate the expected format and
content of your response.

(Example image and answer will be provided below.)
```

**Fig. A6  Few-shot with multi-view and chain-of-thoughts (CoT) reasoning.**

## B. Manufacturing Feature Recognition on Easy, Medium, and Challenging CAD Designs

| Manufacturing Feature Recognition on Easy-Level CAD Images | | | |
|---|---|---|---|
| | Example 4 | Example 5 | Example 6 |
| Isometric View | 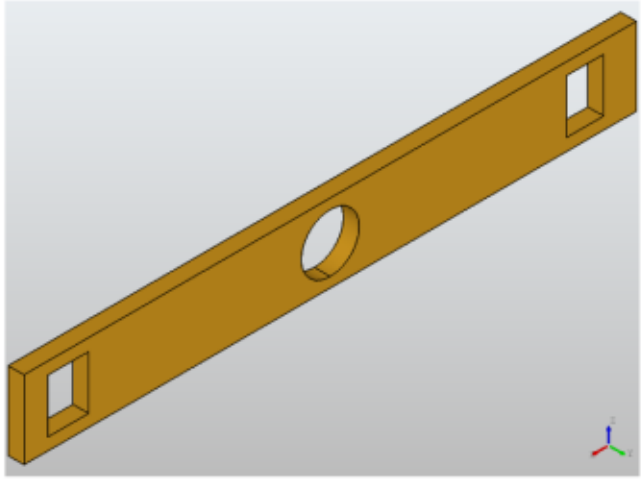 | 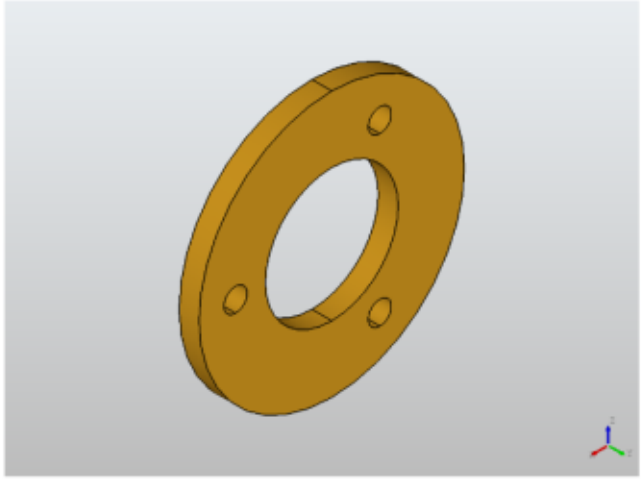 | 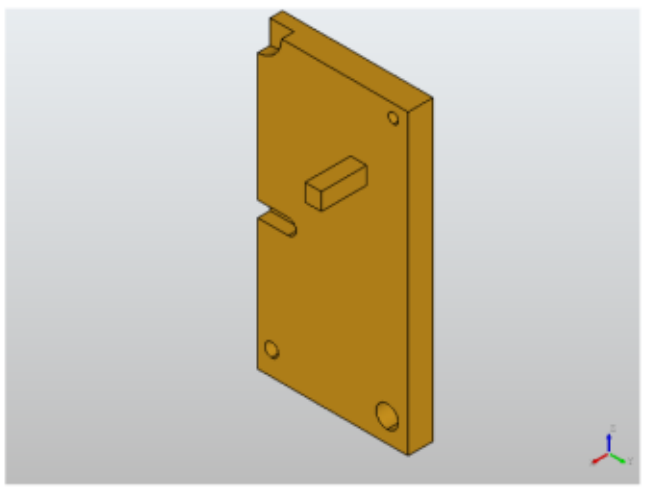 |
| Additional Views | 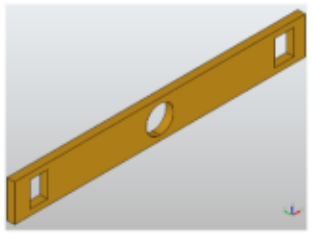 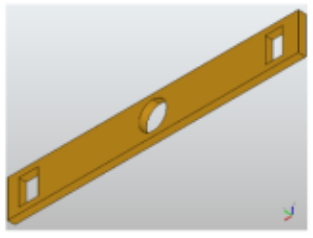 | 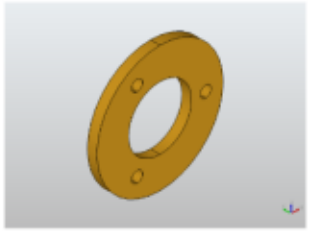 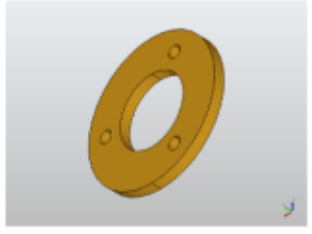 | 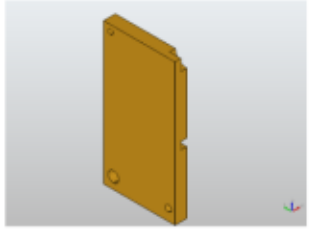 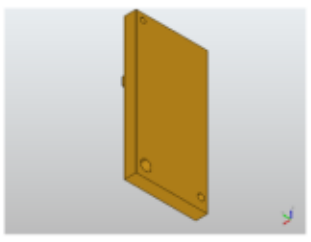 |
| **Ground Truth** (by human expert) | • Slot (through/blind): 2<br>• Hole (through/blind): 1 | • Hole (through/blind): 4 | • Hole (through/blind): 3<br>• Boss: 1<br>• Step (through/blind): 1<br>• Slot (through/blind): 1 |
| **GPT-4o** Exp 6: Few-shot, Multi-view, CoT | • **Hole (through/blind): 1**<br>• **Slot (through/blind): 2** | • **Hole (through/blind): 3** | • **Hole (through/blind): 2**<br>• **Boss: 1**<br>• **Step (through/blind): 1**<br>• **Slot (through/blind): 1** |
| **Claude-3.5** Exp 5: Zero-shot, Multi-view, CoT | • **Hole (through/blind): 1**<br>• **Slot (through/blind): 2** | • **Hole (through/blind): 3**<br>• **Pipe / Tube: 1** | • **Hole (through/blind): 3**<br>• **Boss: 1**<br>• **Slot (through/blind): 2**<br>• **Step (through/blind): 0** |
| **Claude-3** Exp 3: Zero-shot, Multi-view | • **Hole (through/blind): 1**<br>• **Slot (through/blind): 0** | • **Hole (through/blind): 3** | • **Step (through/blind): 1**<br>• **Hole (through/blind): 0**<br>• **Boss: 0**<br>• **Slot (through/blind): 0** |
| **MiniCPM-Llama3** Exp 4: Few-shot, Multi-view | • **Hole (through/blind): 2**<br>• **Slot (through/blind): 0**<br>• **Chamfer/Bevel: 1**<br>• **Step (through/blind): 1**<br>• **Pocket (through/blind): 1** | • **Hole (through/blind): 4**<br>• **Pocket (through/blind): 1**<br>• **Chamfer/Bevel: 2** | • **Hole (through/blind): 2**<br>• **Step (through/blind): 1**<br>• **Boss: 0**<br>• **Slot (through/blind): 0**<br>• **Chamfer/Bevel: 1**<br>• **Pocket (through/blind): 1** |
| **Llava-1.6-mistral-7b** Exp 3: Zero-shot, Multi-view | • **Hole (through/blind): 2**<br>• **Slot (through/blind): 1**<br>• **Step (through/blind): 1**<br>• **Pocket (through/blind): 1**<br>• **Chamfer/Bevel: 1**<br>• **Neck: 1** | • **Hole (through/blind): 2**<br>• **Chamfer/Bevel: 1** | • **Hole (through/blind): 2**<br>• **Slot (through/blind): 1**<br>• **Step (through/blind): 1**<br>• **Boss: 1**<br>• **Pocket (through/blind): 1**<br>• **Chamfer/Bevel: 1**<br>• **Neck: 1** |

*Note: **green** colour represents correct prediction; **red** colour represents false prediction or misprediction; **orange** colour represents hallucination*

**Fig. B1** Manufacturing feature recognition on easy-level CAD images.

| Manufacturing Feature Recognition on Medium-Level CAD Images | | | |
|---|---|---|---|
| | Example 1 | Example 2 | Example 3 |
| Isometric View | 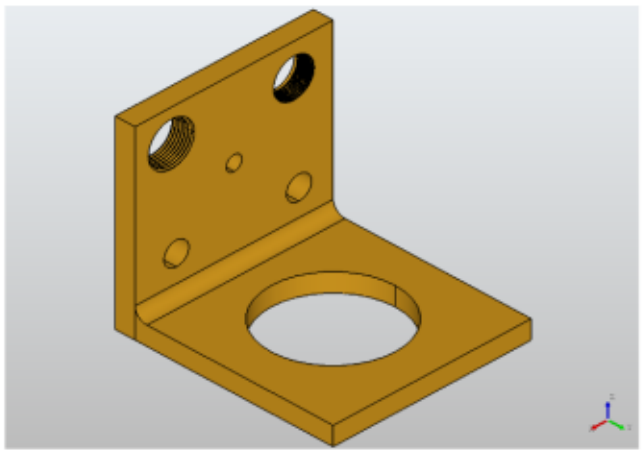 | 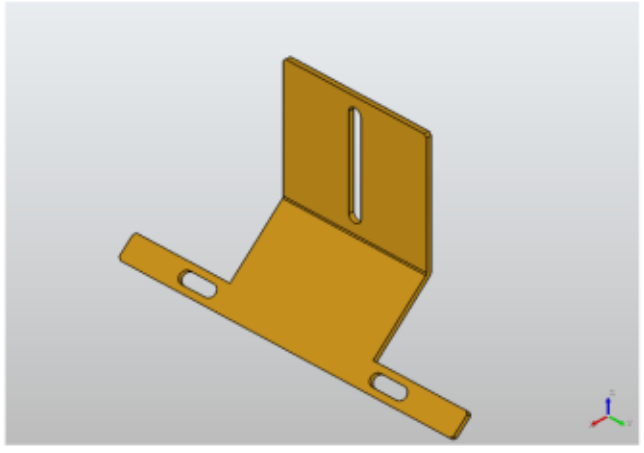 | 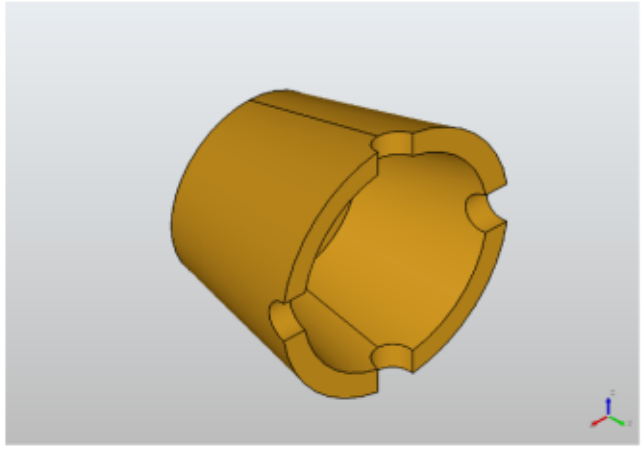 |
| Additional Views | 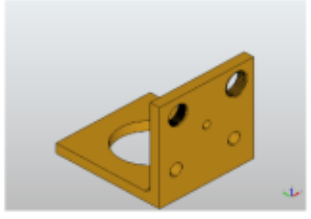 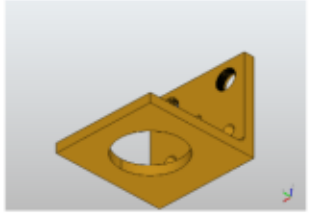 | 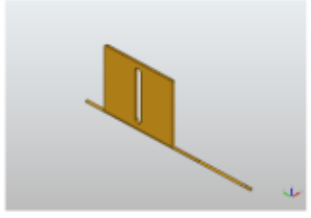 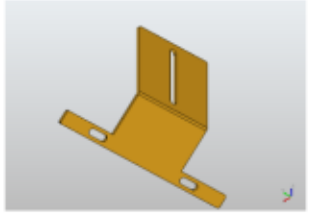 | 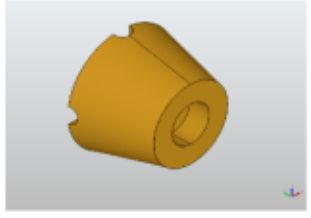 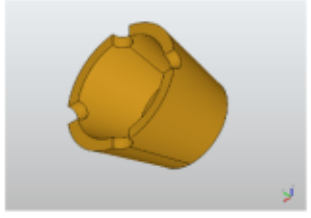 |
| **Ground Truth** (by human expert) | • Hole (through/blind): 4<br>• Fillet/Round: 1<br>• Threaded Feature: 2 | • Slot (through/blind): 3<br>• Sheet Metal: 1 | • Hole (through/blind): 1<br>• Slot (through/blind): 4 |
| **GPT-4o** Exp 6: Few-shot, Multi-view, CoT | • **Hole (through/blind): 4**<br>• **Fillet/Round: 1**<br>• **Threaded Feature: 2** | • **Slot (through/blind): 3**<br>• **Sheet Metal: 1** | • **Hole (through/blind): 1**<br>• **Slot (through/blind): 4** |
| **Claude-3.5** Exp 3: Zero-shot, Multi-view, CoT | • **Hole (through/blind): 5**<br>• **Threaded Feature: 2**<br>• **Fillet/Round: 0**<br>• **Step (through/blind): 1** | • **Slot (through/blind): 3**<br>• **Sheet Metal: 1**<br>• **Step (through/blind): 1** | • **Hole (through/blind): 1**<br>• **Slot (through/blind): 0**<br>• **Chamfer/Bevel: 1** |
| **Claude-3** Exp 5: Zero-shot, Multi-view, CoT | • **Hole (through/blind): 2**<br>• **Fillet/Round: 0**<br>• **Threaded Feature: 0**<br>• **Pocket (through/blind): 1** | • **Slot (through/blind): 0**<br>• **Sheet Metal: 0**<br>• **Hole (through/blind): 1** | • **Hole (through/blind): 0**<br>• **Slot (through/blind): 0**<br>• **Fillet/Round: 2** |
| **MiniCPM-Llama3** Exp 6: Few-shot, Multi-view, CoT | • **Hole (through/blind): 4**<br>• **Fillet/Round: 0**<br>• **Threaded Feature: 0**<br>• **Pocket (through/blind): 1**<br>• **Chamfer/Bevel: 2**<br>• **Freeform Features: 1** | • **Slot (through/blind): 3**<br>• **Sheet Metal: 0**<br>• **Hole (through/blind): 2**<br>• **Step (through/blind): 1**<br>• **Pocket (through/blind): 1**<br>• **Boss: 1**<br>• **Chamfer/Bevel: 1**<br>• **Freeform Features: 1** | • **Hole (through/blind): 1**<br>• **Slot (through/blind): 2**<br>• **Step (through/blind): 1**<br>• **Pocket (through/blind): 1**<br>• **Boss: 1**<br>• **Chamfer/Bevel: 1**<br>• **Freeform Features: 1** |
| **Llava-1.6-mistral-7b** Exp 5: Zero-shot, Multi-view, CoT | • **Hole (through/blind): 2**<br>• **Fillet/Round: 0**<br>• **Threaded Feature: 0**<br>• **Pocket (through/blind): 1**<br>• **Chamfer/Bevel: 1** | • **Slot (through/blind): 1**<br>• **Sheet Metal: 0**<br>• **Hole (through/blind): 2**<br>• **Chamfer/Bevel: 1**<br>• **Fillet/Round: 1** | • **Hole (through/blind): 2**<br>• **Slot (through/blind): 1**<br>• **Pocket (through/blind): 1**<br>• **Chamfer/Bevel: 1**<br>• **Fillet/Round: 1** |

*Note: **green** colour represents correct prediction; **red** colour represents false prediction or misprediction; **orange** colour represents hallucination*

**Fig. B2 Manufacturing feature recognition on medium-level CAD images.**

| Manufacturing Feature Recognition on Difficult-Level CAD Images | | | |
|---|---|---|---|
| | Example 4 | Example 5 | Example 6 |
| Isometric View | 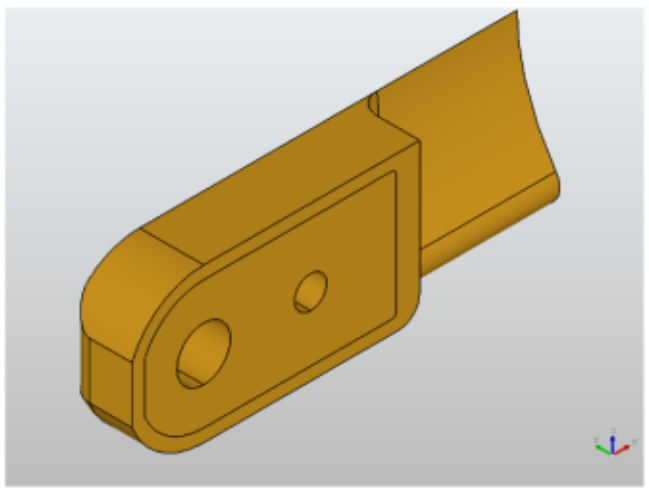 | 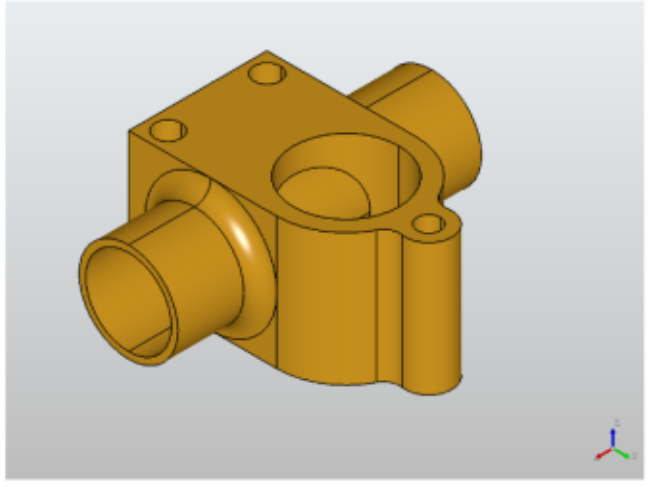 | 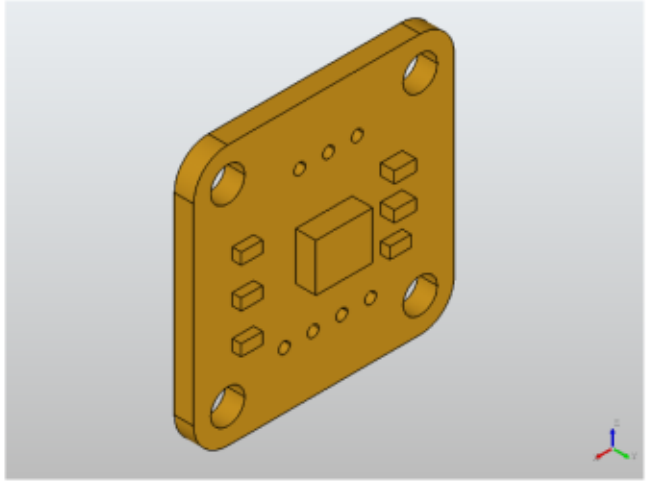 |
| Additional Views | 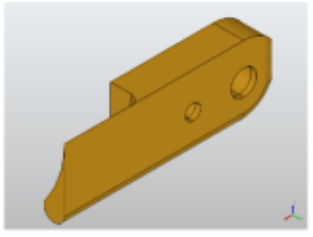 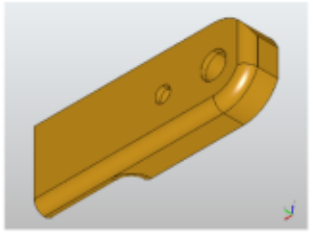 | 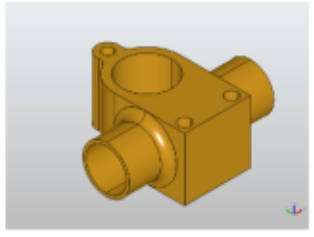 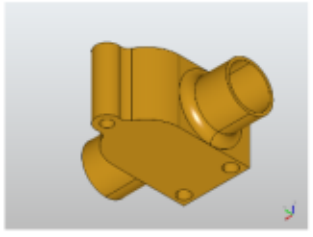 | 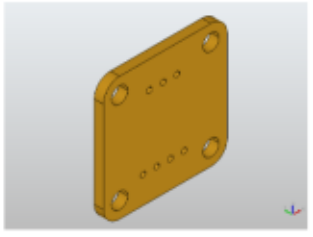 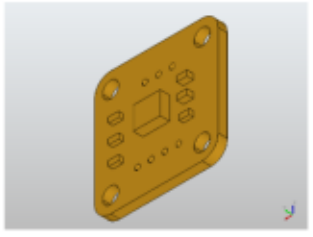 |
| **Ground Truth** (by human expert) | • Hole (through/blind): 2<br>• Fillet/Round: 2<br>• Step (through/blind): 1<br>• Freeform Features: 1 | • Hole (through/blind): 4<br>• Fillet/Round: 2<br>• Boss: 1<br>• Pipe/Tube: 2 | • Hole (through/blind): 11<br>• Fillet/Round: 4<br>• Boss: 7 |
| **Claude-3.5** Exp 3: Zero-shot, Multi-view | • **Hole (through/blind): 2**<br>• **Fillet/Round: 2**<br>• **Step (through/blind): 1**<br>• **Chamfer/Bevel: 1**<br>• **Freeform Features: 0** | • **Hole (through/blind): 4**<br>• **Fillet/Round: 2**<br>• **Boss: 1**<br>• **Pipe/Tube: 2** | • **Hole (through/blind): 4**<br>• **Fillet/Round: 4**<br>• **Boss: 6**<br>• **Pocket (through/blind): 1** |
| **GPT-4o** Exp 5: Zero-shot, Multi-view, CoT | • **Hole (through/blind): 2**<br>• **Step (through/blind): 1**<br>• **Fillet/Round: 2**<br>• **Freeform Features: 0** | • **Hole (through/blind): 3**<br>• **Fillet/Round: 2**<br>• **Pipe/Tube: 2**<br>• **Boss: 0** | • **Hole (through/blind): 4**<br>• **Fillet/Round: 4**<br>• **Boss: 7** |
| **Claude-3** Exp 4: Few-shot, Multi-view | • **Hole (through/blind): 2**<br>• **Fillet/Round: 4**<br>• **Step (through/blind): 0**<br>• **Freeform Features: 0** | • **Hole (through/blind): 4**<br>• **Fillet/Round: 4**<br>• **Pipe/Tube: 0**<br>• **Boss: 0** | • **Hole (through/blind): 6**<br>• **Fillet/Round: 8**<br>• **Boss: 0** |
| **MiniCPM-Llama3** Exp 5: Zero-shot, Multi-view, CoT | • **Hole (through/blind): 4**<br>• **Step (through/blind): 1**<br>• **Chamfer/Bevel: 2**<br>• **Fillet/Round: 0**<br>• **Freeform Features: 0** | • **Hole (through/blind): 4**<br>• **Fillet/Round: 0**<br>• **Boss: 0**<br>• **Pipe/Tube: 0**<br>• **Step (through/blind): 1**<br>• **Chamfer/Bevel: 2** | • **Hole (through/blind): 4**<br>• **Fillet/Round: 0**<br>• **Boss: 0**<br>• **Chamfer/Bevel: 1** |
| **Llava-1.6-mistral-7b** Exp 5: Zero-shot, Multi-view, CoT | • **Hole (through/blind): 2**<br>• **Fillet/Round: 1**<br>• **Chamfer/Bevel: 1**<br>• **Slot (through/blind): 1**<br>• **Freeform Features: 0**<br>• **Step (through/blind): 0** | • **Hole (through/blind): 2**<br>• **Fillet/Round: 1**<br>• **Boss: 0**<br>• **Pipe/Tube: 0**<br>• **Slot (through/blind): 1**<br>• **Poket (through/blind): 1**<br>• **Chamfer/Bevel: 1** | • **Fillet/Round: 1**<br>• **Boss: 1**<br>• **Hole (through/blind): 0** |

*Note:* ***green*** *colour represents correct prediction;* ***red*** *colour represents false prediction or misprediction;* ***orange*** *colour represents hallucination*

**Fig. B3** **Manufacturing feature recognition on challenging CAD images.**